\documentclass[prd,12pt,a4paper,nofootinbib,showpacs,showkeys,tightenlines]{revtex4}
\usepackage[left=2cm,right=2cm,top=2.5cm,bottom=2.5cm]{geometry}

\usepackage[frenchb,english]{babel}
\usepackage[latin1]{inputenc}
\usepackage{enumerate}
\usepackage{amsmath}
\usepackage{amsthm}
\usepackage{amssymb}
\usepackage{graphicx}
\usepackage{floatflt}
\usepackage{fancybox}
\usepackage{stmaryrd}
\usepackage{appendix} 
\usepackage{enumitem}
\usepackage{subfigure}
\usepackage{hyperref} 
\usepackage[b]{esvect}

\usepackage{color}

\renewcommand{\Re}{\textrm{Re}}
\renewcommand{\Im}{\textrm{Im}}

\newcommand{\om}{\omega}
\newcommand{\Om}{\Omega}

\newcommand{\lam}{\lambda}

\newcommand{\eps}{\epsilon}

\newcommand{\s}{{u}}
\newcommand{\+}{{+}}
\renewcommand{\-}{{-}}
\newcommand{\co}{{d}}

\newcommand{\h}{h_B} % Water depth
\newcommand{\omin}{\om_{\rm min}} % Lower threshold 
\newcommand{\oms}{\om_{s}} % Ultra low threshold
\newcommand{\m}{{\rm min}}
\newcommand{\M}{{\rm max}}
\newcommand{\grad}{|\h'|}

\newcommand{\xctp}{x_*} %{x_{\rm tp}^{\mathbb C}} % complex turning point 
\newcommand{\xr}{x_{\mathbb R}} % complex turning point - Real part
\newcommand{\Lvar}{d} % variation length near Fmax
\newcommand{\Fm}{F_{\rm max}}

\newcommand{\sign}{\textrm{sign}}

\newcommand{\p}{\partial}

\newcommand{\be} {\begin{equation}}
\newcommand{\ee} {\end{equation}}
\newcommand{\bsub}{\begin{subequations}}
\newcommand{\esub}{\end{subequations}}
\newcommand{\bea}{\begin{eqnarray}}
\newcommand{\eea}{\end{eqnarray}}
\newcommand{\bi} {\begin{itemize}}
\newcommand{\ei} {\end{itemize}}
\newcommand{\ben} {\begin{enumerate}}
\newcommand{\een} {\end{enumerate}}
\newcommand{\bmat} {\begin{pmatrix}}
\newcommand{\emat} {\end{pmatrix}} 
\newcommand{\bal} {\begin{aligned}}
\newcommand{\eal} {\end{aligned}}
\newcommand{\btab}{\begin{tabular}}
\newcommand{\etab}{\end{tabular}}

\newcommand{\eq}[1]{Eq.~\eqref{#1}}

\begin{document}
\selectlanguage{english}

\title{The imprint of the Hawking effect in subcritical flows}

\author{Antonin Coutant}
\email{antonin.coutant@nottingham.ac.uk}
\affiliation{School of Mathematical Sciences, University of Nottingham, University Park, Nottingham, NG7 2RD, UK}
\author{Silke Weinfurtner}
\email{silke.weinfurtner@nottingham.ac.uk}
\affiliation{School of Mathematical Sciences, University of Nottingham, University Park, Nottingham, NG7 2RD, UK}

\begin{abstract}
We study the propagation of low frequency shallow water waves on a one dimensional flow of varying depth. When taking into account dispersive effects, the linear propagation of long wavelength modes on uneven bottoms excites new solutions of the dispersion relation which possess a much shorter wavelength. The peculiarity is that one of these new solutions has a negative energy. When the flow becomes supercritical, this mode has been shown to be responsible for the (classical) analog of the Hawking effect. For subcritical flows, the production of this mode has been observed numerically and experimentally, but the precise physics governing the scattering remained unclear. In this work, we provide an analytic treatment of this effect in subcritical flows. We analyze the scattering of low frequency waves using a new perturbative series, derived from a generalization of the Bremmer series. We show that the production of short wavelength modes is governed by a complex value of the position: a complex turning point. Using this method, we investigate various flow profiles, and derive the main characteristics of the induced spectrum. 
\end{abstract}


\maketitle

\newpage
\tableofcontents

%%%%%%%%%%%%%%%%%%%%%%%%%%%%%%%%%%%%%%%%%%%%%%%%%%%
%INTRODUCTION
%%%%%%%%%%%%%%%%%%%%%%%%%%%%%%%%%%%%%%%%%%%%%%%%%%%

\newpage
\section{Introduction}
The idea to use fluid flows to mimic the Hawking effect of black holes~\cite{Unruh81}, which allows them to spontaneously emit a thermal radiation, has been intensively studied at the theoretical level~\cite{Jacobson91,Brout95,Barcelo05,Coutant14b}. More recently, several experimental studies have been set up, in various media as diverse as Bose-Einstein condensates~\cite{Steinhauer14,Steinhauer15}, optical fibers~\cite{Belgiorno10,Rubino12} or surface waves, either in water~\cite{Weinfurtner10,Weinfurtner13,Euve14,Euve15}, or in superfluids (therein called ``ripplons''~\cite{Volovik06}). To obtain such a setup, one needs a fluid flow whose velocity crosses the speed of waves. Moreover, because the Hawking effect necessarily involves short wavelength modes, it also necessary to take into account dispersive effects that arise at short distances. One promising possibility is to use surface waves on flowing water~\cite{Schutzhold02,Unruh12}, since their propagation speed is much lower than that of sound waves. When neglecting capillarity and dissipation, surface waves propagate on a one-dimensional homogeneous flow with a frequency $\om$ and a wave number $k$ that obey the dispersion relation~\cite{Johnson,Mei}
\be \label{DispRelFull}
\Om^2 = g k \tanh(\h k), 
\ee
where $\Om = \om - v k$ is the comoving frequency. In this equation, $g$ is the local gravitational field, $v$ the flow velocity, and $\h$ the depth of water. To characterize the flow, it is convenient to introduce the \emph{Froude number}, defined as $F = v/c$, where $c = \sqrt{g \h}$ is the propagation speed of long wavelength waves. If $F>1$ (resp. $F < 1$), the flow is called ``supercritical'' (resp. `subcritical'). The transition from subcritical to supercritical, a ``transcritical flow'', is the analog of a black hole if the flow accelerates, and a white hole if the flow decelerates. Unfortunately, it is experimentally delicate to obtain controllable transcritical flows to study the analog Hawking radiation. One difficulty is caused by the appearance of an undular pattern, or ``undulation'' that deforms the free surface and whose amplitude raises with the Froude number~\cite{Johnson,Coutant12,Coutant13}. Instead, experimental studies have so far focused on flows of high Froude numbers but that stay subcritical~\cite{Weinfurtner10,Weinfurtner13,Euve14,Euve15}. In these cases, due to dispersive effects one still observes the production of the negative norm mode responsible for the Hawking effect in transcritical flows. However, theoretical treatments on the analog Hawking effect have mostly focused on transcritical flows, and it is therefore presently unclear what governs the spectrum of this negative norm mode production for subcritical flows. Recent numerical works~\cite{Finazzi11,Michel14,Michel15} have indicated that the spectrum is in general quite different when the flow is subcritical with respect to the transcritical case. 

In this work, we provide an analytical characterization of the low frequency scattering when the flow is inhomogeneous, i.e. when $c$, $v$, and $\h$ depend on the position $x$. For this, we developed a new mathematical approach, based on a generalization of the Bremmer series~\cite{Bremmer51,Landauer51,Berry72}. This allows us to obtain a perturbative expansion of the various scattering coefficients in gradients of the background. More precisely, the ``small parameter'' of the expansion will be played by the variation of the height of the flow, i.e. $\grad$. The first order treatment shows that the values of the scattering coefficients are mainly governed by \emph{complex turning points}. At very low frequencies, all the turning points reduce to a single \emph{complex horizon}, which is the locus where the Froude number $F(x)$ reaches 1 if it is analytically continued to complex positions $x$. Although we focus in this work on water waves, we believe that our conclusions are still valid for subcritical flows (or what replaces it) in other analog gravity systems where dispersion decreases the velocity at short wavelengths, e.g., optical fibers~\cite{Belgiorno10,Rubino12}, or sound in a duct~\cite{Auregan15}. The paper is divided as follows. In the first section we present the setup and the wave equation. In the second one, we derive the perturbative series, and show that the first order is given in terms of contour integrals involving the complex turning points. In the last section, we apply this general framework to specific flow examples, discuss the various regimes and the relevant physics for present experiments. In Appendices, we provide very general proofs, preparing our results for further extensions.

\section{The settings}
\subsection{Surface wave equation}
We shall consider the propagation of water waves in the so-called ``weak dispersive regime''. In this regime, the dispersion relation \eqref{DispRelFull} is approximated by the first two terms of the low $k$ expansion, i.e. $k \tanh(\h k) \sim \h k^2 - \h^3 k^4/3$. As mentioned in the introduction, it is necessary to take into account dispersive effects since the scattering processes we are interested in involve short wavelength modes. When the maximum value of the Froude number $\Fm$ is close to 1, a case referred to as `near critical flows', the weak dispersive regime provides a good approximation of the scattering coefficient. However, we believe that even for flows that are not near critical, the \emph{qualitative} features we describe will be very similar. In the weak dispersive regime, gravity waves are described by the action~\cite{Coutant13}~\footnote{Notice that there is a small difference with~\cite{Coutant13}. There is a change of ordering between $\p_x$ and $\h(x)$ in the last term. The action of~\cite{Coutant13} is more accurate, but our choice makes a couple of equations simpler. Moreover, the difference will only show up at third order in the small parameter $\grad$, and therefore is irrelevant for our present purpose.} 
\be \label{action}
\mathcal S = \frac12 \int \left[(\p_t \phi + v(x) \p_x \phi)^2 - c^2(x) (\p_x \phi)^2 + \frac{g \h^3(x)}{3} (\p_x^2 \phi)^2\right] dtdx. 
\ee
In the following, $v$ is assumed to be positive, so that water flows from left to right. The field $\phi$ encodes the fluctuations of the velocity potential of the flow at the surface. It is directly related to the change of height of the free surface. In the presence of a (linear) wave, the water depth becomes $\h(x) + \delta h(t,x)$. The surface elevation $\delta h$ is then given by  
\be
\delta h(t,x) = -\frac1g (\p_t + v\p_x) \phi, 
\ee
where this follows from Bernouilli's equation~\cite{Unruh12,Coutant13}. Minimizing the action \eqref{action} gives us the equation of motion for the field 
\be \label{TEoM}
(\p_t + \p_x v)(\p_t + v \p_x) \phi - \p_x c^2 \p_x \phi - \frac{g}3 \p_x^2 \h^3 \p_x^2 \phi = 0. 
\ee
We assume that the background flow is stationary, and therefore, we look for solutions of \eq{TEoM} at fixed frequency, i.e. of the form $\phi = \Re\left(\phi_\om(x) e^{-i \om t}\right)$, where $\phi_\om(x)$ is a \emph{complex} stationary mode. Since time-dependent solutions are obtained by taking the real part, it is enough to work with $\om > 0$. The modes $\phi_\om(x)$ satisfy the equation 
\be \label{EoM}
(\om + i \p_x v)(\om + i v \p_x) \phi_\om = - \p_x c^2 \p_x \phi_\om - \frac{g}3 \p_x^2 \h^3 \p_x^2 \phi_\om .
\ee
Before trying to solve this equation, it is useful to analyze its main properties. As it is derived from an action, it possesses a canonically conserved norm, given by 
\be \label{KGnorm}
(\phi | \phi) = \int \rho[\phi] dx = - \int \Im \Big(\phi^* (\p_t + v \p_x) \phi \Big) dx. 
\ee
This norm plays a crucial role in the characterization of the scattering. For positive frequency modes, the sign of the norm coincides with that of the energy. As we shall see, due to dispersion, the system possesses \emph{negative energy modes}, or equivalently, modes with a negative norm \eqref{KGnorm}. This is characteristic of unstable flows~\cite{Fabrikant}. The generation of a negative norm mode by sending a positive norm one is referred to as ``anomalous scattering''. For transcritical flows, this scattering (in the smooth limit) is the classical analog of the Hawking effect~\footnote{We refer the reader to the literature on the Hawking effect in water waves~\cite{Schutzhold02,Unruh12,Rousseaux13} and in particular~\cite{Coutant13}, where the role of the scalar product and the energy is discussed with care, see Appendix B.}. For subcritical flows, it is still present, but it was so far unclear what governs the spectrum, i.e., the values of the scattering coefficients for various frequencies. As we shall deal exclusively with stationary modes, it is more convenient to work with the conserved current rather than the norm \eqref{KGnorm}. Because of the dispersive term, this current is not the standard Klein-Gordon current, but has a more complicated form~\cite{Richartz12}. Starting from the action \eqref{action}, it reads 
\be \label{Current_Eq}
J[\phi_\om] = \Im \Big( i \om v \phi_\om^* \phi_\om + (c^2 - v^2) \phi_\om^* \p_x \phi_\om + \frac{g}3 \phi_\om^* \p_x \h^3 \p_x^2 \phi_\om - \frac{g\h^3}3 \p_x \phi_\om^* \p_x^2 \phi_\om \Big). 
\ee
For any mode solution of \eqref{EoM}, the current is (exactly) conserved, i.e. $\p_x J = 0$. This current represents the amount of \emph{norm} that is transported by a mode. Its conservation is of course equivalent to that of the norm \eqref{KGnorm}. This can be directly seen from the identity $\p_t \rho + \p_x J = 0$ (see App.~\ref{Current_App}), which follows from the application of the Noether theorem to \eq{action}. Notice also that for $\om > 0$, $\om J[\phi_\om]$ is the \emph{energy current}~\cite{Coutant13}. 

\begin{figure}[!ht]
\begin{center}
\includegraphics[width=0.8\columnwidth]{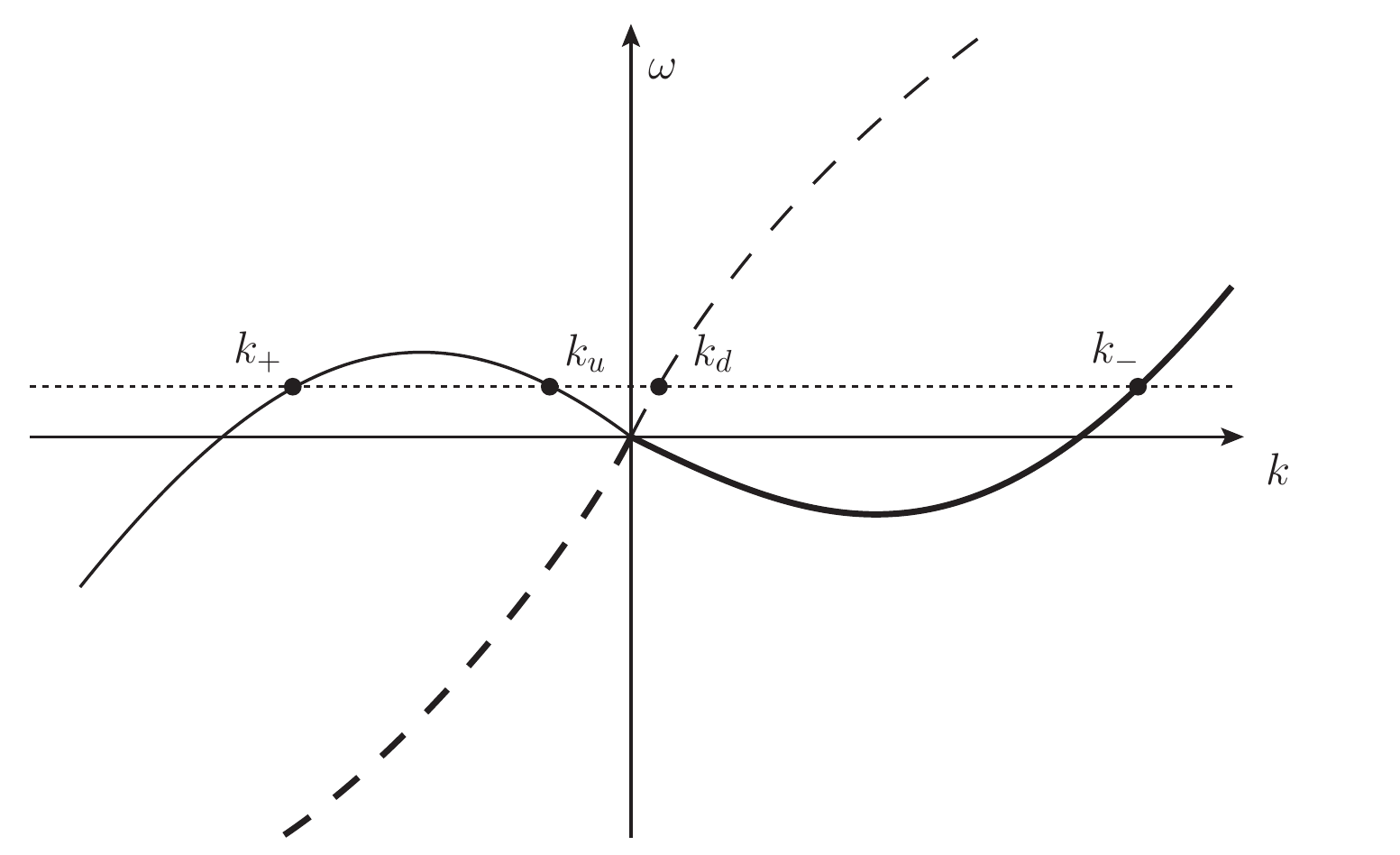}
\end{center}
\caption{Graphical resolution of the dispersion relation \eqref{Disp_rel}. The continuous line is the counter-propagating branch (upstream), while the dashed line is the co-propagating branch (downstream). The bold lines indicates $\Om(k) < 0$, where $\Om$ is defined after \eq{DispRelFull}. 
}
\label{HJ_Fig} 
\end{figure}

When the background flow is homogeneous, i.e. $c$, $v$, and $\h$ are constant, the solutions are superpositions of plane waves $e^{i k_\om x}$. Here, $k_\om$ is the wave number, or momentum, and satisfies the dispersion relation 
\be \label{Disp_rel} 
(\om - v k_\om)^2 = c^2 k_\om^2 - \frac{g \h^3 k_\om^4}{3} .
\ee
Below a certain threshold frequency $\om_{\rm crit}$, this equation possesses 4 distinct roots (see Fig.~\ref{HJ_Fig}). Two of them have long wavelengths, while the two others have short wavelengths. The first two are the usual left-mover (noted $k_\s$, as it moves against the flow, i.e. ``upstream'') and right-mover (noted $k_\co$, for ``downstream''). The two other roots, which are absent when the flow velocity vanishes, are due to both the nonzero flow and dispersion. One of them, $k_\-$, has a positive value but a \emph{negative norm}. The other have a negative value and a \emph{positive norm}, and is denoted $k_\+$. (The index refers to the sign of the norm.) The negative norm mode described by $k_\-$ will play a crucial role in the following. 

The aim of this work is to study, when the flow becomes inhomogeneous, how these four modes mix. In particular, we shall see how the propagation of a long wavelength left-mover $k_\s$ generates the short wavelength modes, as was experimentally observed in~\cite{Weinfurtner10,Euve15}. For flows that become critical, this generation is the classical analog of the Hawking effect. For subcritical flows, such a mode conversion still exist, but the law governing the scattering coefficients was so far not known analytically. This is what we aim at characterizing.

\subsection{Characterization of the background flow}
\label{Flow_Sec}
\begin{figure}[!ht]
\begin{center}
\includegraphics[width=0.8\columnwidth,trim=0 3cm 0 0,clip]{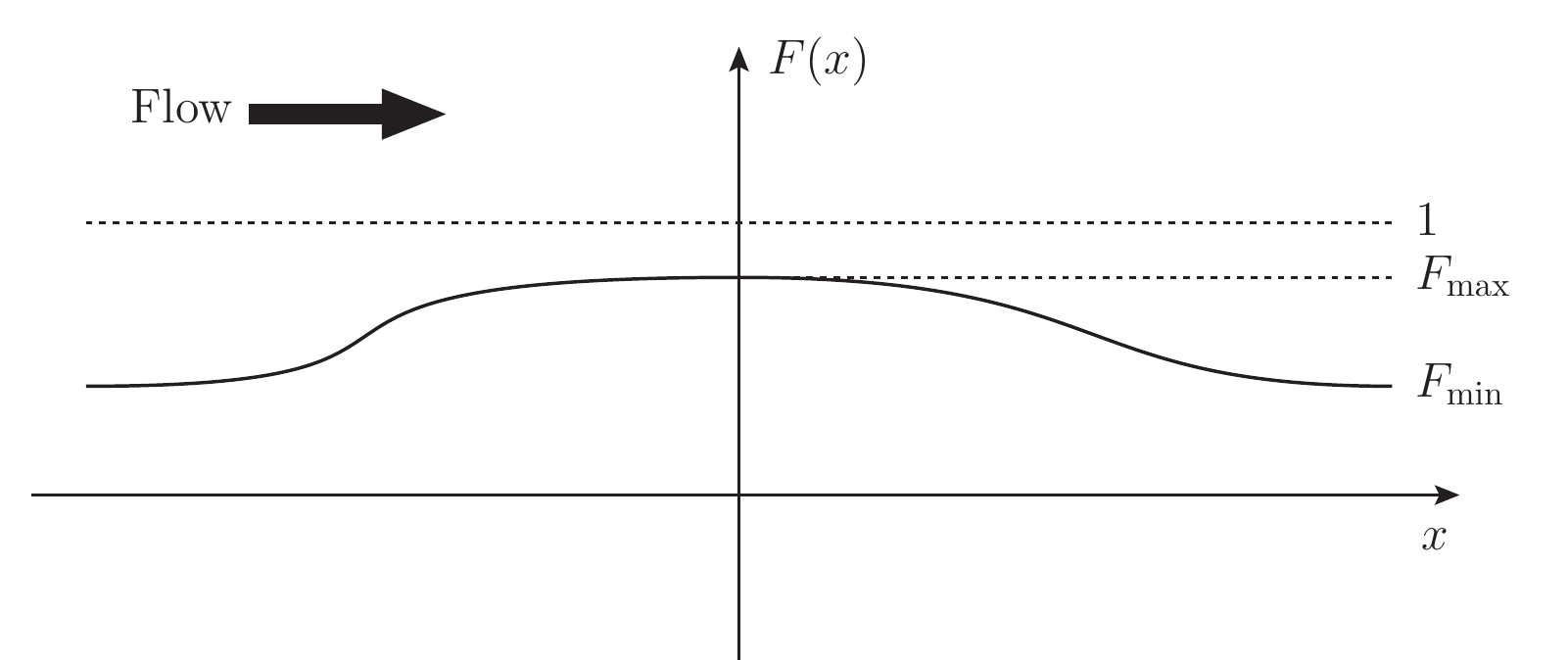}
\end{center}
\caption{Shape of the Froude number as a function of $x$. The bold arrow indicates the direction of the flow.
}
\label{Fr_Fig} 
\end{figure}

We assume that the fluid flows over a smooth obstacle. The height of fluid $\h(x)$ varies monotonically from an asymptotic value on the left side, to a minimum value $h_\m$ and increases again to a constant value on the right side. When the obstacle is smooth enough, the (unperturbed) free surface stays approximately flat, and in this case, the other background quantities are directly deduced from the height by the relations 
\bsub \label{IdealFlow} \bea
c(x) &=& \sqrt{g \h(x)} , \\
v(x) &=& \frac{q}{\h(x)} , 
\eea \esub
where $g$ is the local gravitational acceleration, and $q$ the (conserved) flow rate (water flux per unit width, expressed in $\mathrm{m^2 \cdot s^{-1}}$). As we shall see, the most relevant quantity to describe the flow is the local value of the Froude number 
\be
F(x) \equiv \frac{v(x)}{c(x)} = \frac{q}{g^{1/2} \h(x)^{3/2}}, 
\ee
where the second equality is satisfied when \eqref{IdealFlow} is. In realistic flows, curvature effects (but also dissipation) will deform the free surface, and the relation between these  quantities becomes more intricate~\footnote{In ~\cite{Unruh12}, a different wave equation was proposed, which takes into account effects from the curvature of the free surface. Later, the link between this equation and the more familiar \eqref{TEoM} was established in~\cite{Coutant13}. In particular, it was shown that the corrections due to curvature can be implemented by using the same equation \eqref{TEoM}, but where $c$, $v$, and $\h$ are related by a more intricate relation than \eqref{IdealFlow}, see Eq.~(3) therein.}. Unless otherwise specified, we will treat the three functions $v$, $c$, and $\h$ as independent, thereby leaving the possibility to include corrections to \eq{IdealFlow}. However, to keep control on the various approximations, we assume that the gradients, in units of the dispersive scale, are essentially of the same order, i.e., that $|\h v'/v|$ and $|\h c'/c|$ are of the same order as $\grad$, and we refer to the ``smooth limit'' as $\grad \ll 1$. This is automatically the case if the three functions are related by \eq{IdealFlow}. Far from the obstacle, we assume that the background quantities $v$, $c$, and $\h$ are constant. Over this obstacle, the flow velocity $v$ increases to a maximum $v_\M$ while the wave speed decreases to a minimum $c_\m$. At the top of the obstacle, the Froude number reaches its maximum $\Fm$ (see Fig.~\ref{Fr_Fig}). The assumptions we make are in practice nontrivial. First it assumes that no turbulence is formed by the flow close to the free surface. While this is reasonable for subcritical flows, we also assumed that no undulation appears at the free surface. When increasing the Froude number, even below 1, such an undulation is more likely to form, as observed in~\cite{Weinfurtner10,Euve15}. We believe that the presence of such an undulation could be treated by our framework (see the remark of footnote~\ref{Undulation_ftn}), but the computations will be more involved. We feel that such an analysis goes beyond the scope of the present paper. 

\subsection{The WKB approximation}
\label{WKB_Sec}
When $v(x)$, $c(x)$, and $\h(x)$ vary, plane waves are no longer solutions of \eq{EoM}. In the limit of a very smooth background $\grad \to 0$, solutions of \eq{EoM} are given by WKB modes, i.e., locally plane waves characterized by a local momentum $k_\om(x)$. This local momentum is a solution of the Hamilton-Jacobi equation, which is nothing else than the dispersion relation \eqref{Disp_rel} in an inhomogeneous background 
\be \label{HJ}  
(\om - v(x) k_\om)^2 = c(x)^2 k_\om^2 - \frac{g}{3} \h^3(x) k_\om^4.
\ee
Each solution of $k_j$ of this equation depends on both $\om$ and $x$. To lighten the notations, we shall drop this dependance when unnecessary.  Throughout this paper, we also assume that the 4 roots are real and distinct for all $x$. Since the flow stays subcritical all along, i.e. $\Fm < 1$, this is realized below a threshold frequency $\omin = \min_x(\om_{\rm crit}(x))$. For near critical flows, the value of this frequency reads 
\be \label{omin_eq}
\omin \sim \frac{c_\m}{3 h_\m} \big(1-\Fm^2 \big)^{3/2} .
\ee
For $\om < \omin$, a WKB mode is then given by 
\be \label{WKBsup}
\varphi_j(x) = A_j e^{i \int k_j(x') dx'}, 
\ee
where the subscript $j$ indicates the corresponding Hamilton-Jacobi root, i.e. $j \in \{\s, \+, \-, \co\}$. A careful analysis (see e.g. Appendix A of~\cite{Coutant11} or Appendix~\ref{Math_App} of this work) shows that in the limit of smooth backgrounds, the amplitude simply reads 
\be \label{WKBampl}
A_j = \frac1{\sqrt{|\Om(k_j) v_g(k_j)|}}, 
\ee
where $v_g$ is the group velocity of the corresponding mode, and $\Om$ its co-moving frequency, defined after \eq{DispRelFull}. Moreover, at the level of the WKB approximation, the current $J$ is easy to compute and one sees that 
\be
J[\varphi_j] = \pm 1 = \sign \left(\Om(k_j) v_g(k_j)\right). 
\ee
Hence, the WKB amplitude \eqref{WKBampl} normalizes the current of a WKB mode to $\pm 1$~\cite{Richartz12}. This property of WKB modes comes from the fact that the current $J$ is an \emph{adiabatic invariant} of the problem~\cite{Buhler}. Unfortunately, the WKB approximation precisely consists in neglecting the mode mixing, which is what we are after. To overcome this problem, we first notice that for $\om < \omin$, no crossing occurs, i.e. the 4 roots of \eqref{HJ} are distinct for all $x$. Therefore, the 4 WKB modes in \eqref{WKBsup} are perfectly well-defined functions of $x$. Instead of using them as approximate solutions of the wave equation, we shall use them as a new \emph{basis} to represent \emph{exact} solutions of the wave equation \eqref{EoM}. This allows us to recast the wave equation in an equivalent form, adapted to a perturbative expansion of the scattering coefficients in the background gradients.

\section{Beyond WKB: the local scattering coefficients}
\subsection{The Bremmer representation}
\label{Bremmer_Sec}
The idea to use the WKB modes as a basis has been widely studied and used for second order differential equations, where it is called the Bremmer series. It has a wide range of applications, from scattering theory of the Schrödinger equation~\cite{Berry72}, or wave propagation in inhomogeneous media~\cite{Bellman}, %A reference added
to particle production in early cosmology~\cite{Massar97,Winitzki05}. Here however, we must use an extended version of this method, as the problem is intrinsically higher order. As explained in the previous section, the $S$-matrix is $4\times 4$. In Appendix~\ref{Math_App}, we present detailed proofs of how to extend the Bremmer series for higher order equations. In this section, we present the method without technical calculation, in order to focus on the physics and the significance of this new representation. The key idea is to write general, exact solutions of \eq{EoM} as superpositions of WKB waves, where the various amplitudes are $x$-dependent, i.e. 
\be \label{Adiab0}
\phi(x) = A_\s(x) e^{i \int k_\s(x') dx'} + A_\+(x) e^{i \int k_\+(x') dx'} + A_\-(x) e^{i \int k_\-(x') dx'} + A_\co(x) e^{i \int k_\co(x') dx'} .
\ee
At this level, the function $A_j(x)$ are unspecified functions, and are \emph{not} given by \eqref{WKBampl}. Since this introduces 4 unknown functions, instead of 1, we impose 3 extra conditions. The idea is to decompose also the first, second and third derivatives of $\phi(x)$ on the WKB basis, and the forth derivative will then be given by the equation of motion \eqref{EoM}. Explicitly, we assume, in addition to \eqref{Adiab0}, 
\bsub \bea 
-i \p_x \phi &=& k_\s A_\s e^{i \int k_\s(x') dx'} + k_\+ A_\+ e^{i \int k_\+(x') dx'} + k_\- A_\- e^{i \int k_\-(x') dx'} + k_\co A_\co e^{i \int k_\co(x') dx'} , \qquad \label{Adiab1} \\
- \p_x^2 \phi &=& k_\s^2 A_\s e^{i \int k_\s(x') dx'} + k_\+^2 A_\+ e^{i \int k_\+(x') dx'} + k_\-^2 A_\- e^{i \int k_\-(x') dx'} + k_\co^2 A_\co e^{i \int k_\co(x') dx'} , \label{Adiab2} \\
i \p_x^3 \phi &=& k_\s^3 A_\s e^{i \int k_\s(x') dx'} + k_\+^3 A_\+ e^{i \int k_\+(x') dx'} + k_\-^3 A_\- e^{i \int k_\-(x') dx'} + k_\co^3 A_\co e^{i \int k_\co(x') dx'} . \label{Adiab3}
\eea \esub 
Using these 3 conditions and the main \emph{ansatz} \eqref{Adiab0}, we show that the knowledge of $\phi(x)$ is \emph{equivalent} to the knowledge of $A_\s(x)$, $A_\+(x)$, $A_\-(x)$ and $A_\co(x)$. The 4 equations combine to give the single matrix equation 
\be \label{Bremmer_repr}
\bmat \phi(x) \\ -i \p_x \phi(x) \\ -\p_x^2 \phi(x) \\ i \p_x^3 \phi \emat = V \cdot \bmat A_\s(x) e^{i \int k_\s(x') dx'} \\ A_\+(x) e^{i \int k_\+(x') dx'} \\ A_\-(x) e^{i \int k_\-(x') dx'} \\ A_\co(x) e^{i \int k_\co(x') dx'} \emat, 
\ee
where $V$ is the Vandermonde matrix of the 4 roots $k_\s$, $k_\+$, $k_\-$, and $k_\co$, i.e. 
\be \label{VdM}
V = \bmat 1 & 1 & 1 & 1 \\ k_\s & k_\+ & k_\- & k_\co \\ k_\s^2 & k_\+^2 & k_\-^2 & k_\co^2 \\ k_\s^3 & k_\+^3 & k_\-^3 & k_\co^3 \emat. 
\ee
Because the three roots are distinct, $\det(V) \neq 0$, and hence, the relation between $\phi$ and its derivatives and $(A_\s, A_\+, A_\-, A_\co)$ is one-to-one. Therefore, the wave equation \eqref{EoM} can now be entirely recast in an equivalent equation for the local amplitudes $A_j(x)$. To obtain the equation satisfied by the local amplitudes $A_j(x)$, we plug the \emph{ansatz} \eqref{Bremmer_repr} in the wave equation \eqref{EoM}. Since the first three derivatives of $\phi$ are given by \eqref{Bremmer_repr}, we are left with a first order equation on the four amplitudes $(A_\s, A_\+, A_\-, A_\co)$ (see App.~\ref{Bremmer_App}). This equation has the form 
\be \label{Ampl_Eq}
\p_x A_j  = \widetilde{\mathcal M}_{jj}(x) A_j + \sum_{\ell \neq j} \widetilde{\mathcal M}_{j \ell}(x) e^{i \int (k_\ell(x') - k_j(x')) dx'} A_\ell . 
\ee
The off-diagonal elements of $\widetilde{\mathcal M}$ are easy to interpret: they give the coupling between the different WKB branches, due to the varying background. Those are responsible for the nontrivial scattering. On the other hand, the diagonal terms of \eq{Ampl_Eq} represent the adiabatic evolution of the amplitudes $A_j(x)$. To further simplify the equation, we can integrate these diagonal terms by working with normalized amplitudes. For this, we define 
\be
A_j(x) = a_j(x) \mathcal{N}_j(x), 
\ee
where $\mathcal{N}_j$ is chosen so that the first term of \eqref{Ampl_Eq} disappears. This gives a first order equation on $\mathcal{N}_j$, which directly integrate as $\mathcal{N}_j = \exp\left(\int^x \widetilde{\mathcal M}_{jj}(x') dx'\right)$. As we show in App.~\ref{Mmat_App} and~\ref{N4_App}, this leads to %this equation can be directly integrated, and one finds 
\be
\mathcal{N}_j = \frac1{\sqrt{|\Om(k_j) v_g(k_j)|}}. 
\ee
We recognize here nothing else than the WKB amplitude given in \eq{WKBampl}. This is not a surprise, as $\mathcal{N}_j$ gives the adiabatic evolution of the amplitudes. We shall refer to these new coefficients $a_j(x)$ as the \emph{local scattering coefficients}. At the level of the WKB approximation, they are constant. When the background varies, these coefficients becomes non constant, meaning that the propagation of one mode excites the other ones, leading to nontrivial asymptotic scattering coefficients. These coefficients are governed by a first order equation, directly obtained from \eqref{Ampl_Eq}, and which reads 
\be \label{Scat_Eq}
\p_x a_j = \sum_{\ell \neq j} \mathcal M_{j \ell}(x) e^{i \int (k_\ell(x') - k_j(x')) dx'} a_\ell .
\ee
This equation possesses several key features, that we now wish to underline. First, this equation is strictly equivalent to the original equation \eqref{EoM}. No approximation have been used so far, but this rewriting is very adapted to a perturbative resolution. Second, the coupling coefficients $\mathcal M_{j \ell}$ are proportional to derivatives of the background. In the limit $\grad \ll 1$, they are small and have a slowly varying phase (see App.~\ref{Mmat_App} for their exact expressions). Because of this, the coefficients $a_j$ mainly couple through the change of their WKB phases $e^{i \int (k_\ell(x') - k_j(x')) dx'}$. This structure implies that the scattering will become significant when this phase difference satisfies a resonance condition (see next section). The last key property of \eq{Scat_Eq} is obtained when computing the conserved current \eqref{Current_Eq} in terms of the local scattering coefficients. Since $J$ involves only the first three derivatives of $\phi$, the \emph{ansatz} \eqref{Bremmer_repr} guarantees that the computation of $J$ is identical as in the case of plane waves. After some effort (shown in App.~\ref{Current_App}), we show that 
\be \label{Jconserv}
J = - |a_\s(x)|^2 + |a_\+(x)|^2 - |a_\-(x)|^2 + |a_\co(x)|^2 = \mathrm{const}. 
\ee
Once again, this equation is \emph{exact}. It guarantees that the scattering governed by \eq{Scat_Eq} conserves the norm of \eq{KGnorm}. Also, from \eq{Scat_Eq}, the conservation of the current implies that the matrix $\mathcal M$ has some symmetric/antisymmetric properties, something that is not transparent from their explicit expressions (given in App.~\ref{Mmat_App}).

\subsection{The complex turning points}
\label{tp_Sec}
\begin{figure}[!ht]
\begin{center}
\includegraphics[width=0.8\columnwidth]{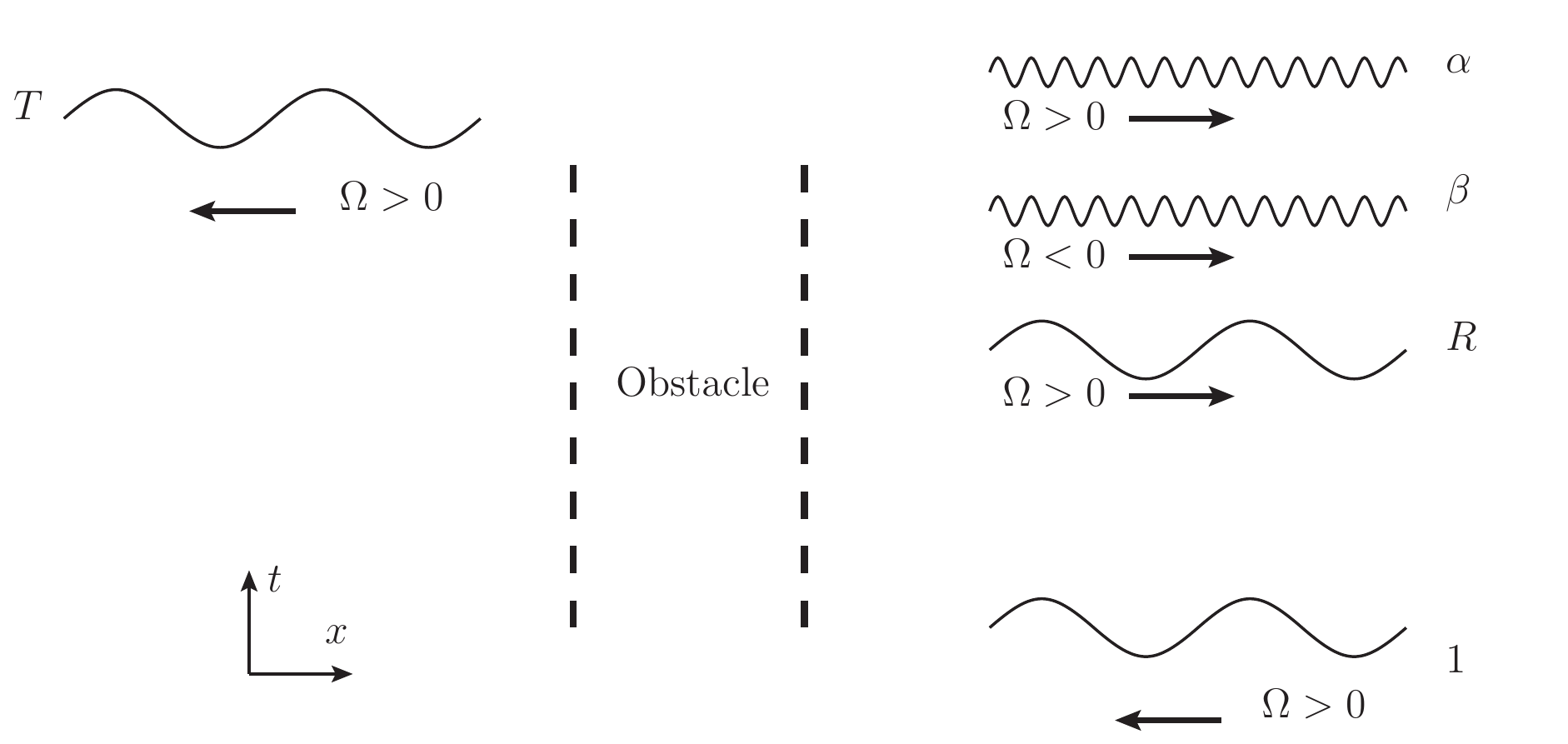}
\end{center}
\caption{Space-time picture of the scattering. Far from the obstacle, the background flow becomes constant and the solution reduces to a superposition of plane waves. The asymptotic values of the local scattering coefficients give the (global) scattering coefficients, i.e. $a_\s(-\infty) = T$, $a_\+(+\infty) = \alpha$, $a_{\-}(+\infty) = \beta$, and $a_\co(+\infty) = R$. 
}
\label{Scat_Fig} 
\end{figure}

We now turn to the evaluation of the scattering coefficients in the limit of smooth backgrounds $\grad \ll 1$. Since 4 modes exist on both sides, there are 4 incoming legs and 4 outgoing ones, and the complete $S$-matrix is $4\times 4$. In a quantum mechanical language, the scattering coefficients can be seen as transition amplitudes for a mode transition $k_\ell \to k_j$. These transitions can then be estimated in perturbation theory of \eq{Scat_Eq}, i.e. in an expansion in the matrix elements $\mathcal M_{\ell j}$, which are small in smooth backgrounds $\grad \ll 1$. For the present purpose, we shall consider only one specific scattering mode, but our results easily extends to the others. We consider a long wavelength mode coming in from the right, which means that $a_\s(+\infty) = 1$, and $a_\+(-\infty) = a_\-(-\infty) = a_\co(-\infty) =  0$ (see Fig.~\ref{Scat_Fig}). This fixes half of the asymptotic values of the local scattering coefficients. The other half gives the scattering coefficients (see Fig.~\ref{Scat_Fig}). $T$ and $R$ are the transmission and reflection coefficients between the two long wavelength modes, while $\alpha$ and $\beta$ describe the generation of the short wavelength modes. Using \eqref{Jconserv}, the conservation of the current imposes the following relation between the scattering coefficients 
\be \label{Unitarity}
|T|^2 + |R|^2 + |\alpha|^2 - |\beta|^2 = 1. 
\ee
We see that the coefficient $\beta$ contributes with the unusual sign. 

\begin{figure}[!ht]
\begin{center}
\includegraphics[width=0.8\columnwidth]{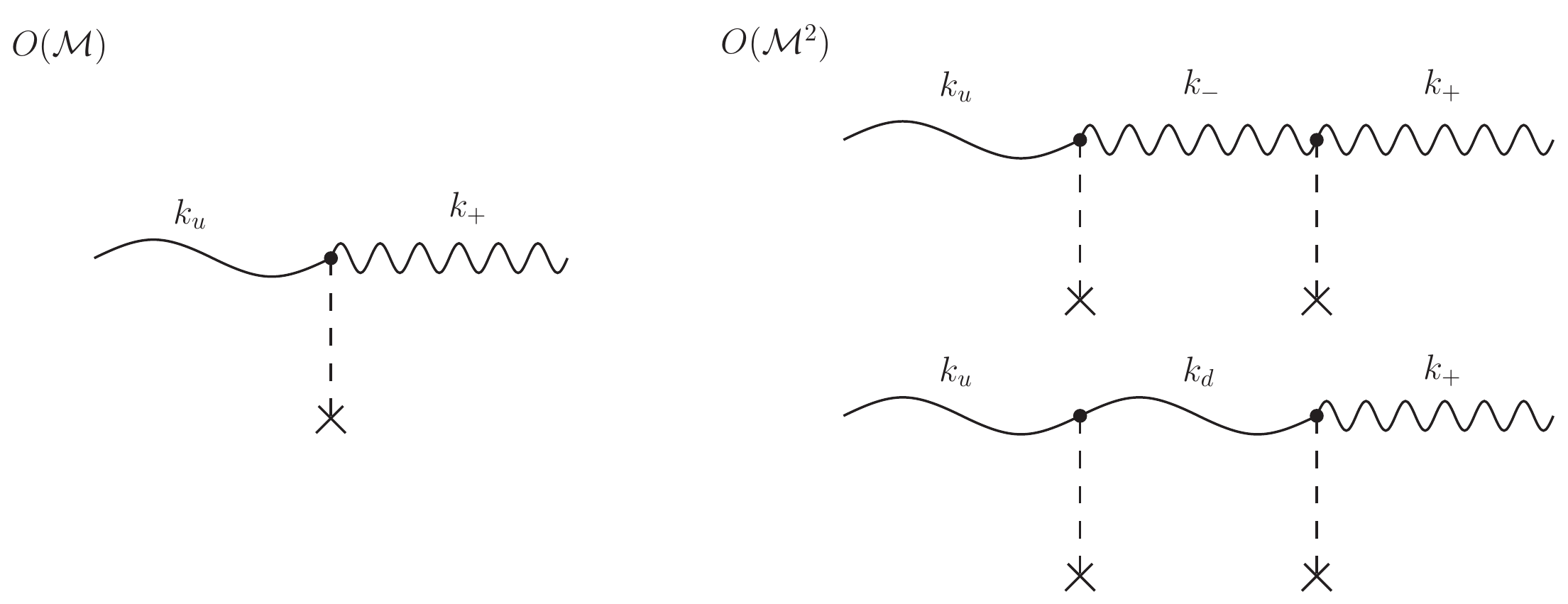}
\end{center}
\caption{Diagrammatic representation of the perturbative resolution of \eq{Scat_Eq} for the transition $k_\s \to k_\+$, i.e., the coefficient $\alpha$. The dashed lines symbolize the fact that the various modes interact through the background, i.e. through $v'$, $c'$, and $\h'$. The expressions we provide in this section~\ref{tp_Sec} are at first order $O(\mathcal M)$. 
}
\label{GraphO1_Fig} 
\end{figure}

At zeroth order in $\mathcal M$, $a_\s(x) \sim 1$, while the other $a_j(x)$ vanish. When inserting this on the right hand side of \eqref{Scat_Eq}, and integrating from $-\infty$ to $+\infty$, we obtain the first order expression in $\mathcal M$ of the scattering coefficients. At this order, $T$ is 1, while the other three are given by an integral expression. To start, we focus on the computation of $\alpha$. By solving \eq{Scat_Eq} at leading order, only the coefficient $\mathcal M_{\+ \s}$ contributes and we have 
\be \label{AlphaInt}
\alpha \sim \int_{-\infty}^{+\infty} \mathcal M_{\+ \s}(x) e^{i \int \left(k_\s(x') - k_\+(x')\right) dx'} dx. 
\ee
This gives the first order expression for the coefficient $\alpha$. It is possible, starting from \eq{Scat_Eq} to derive an expression for $\alpha$ at any order in $\mathcal M$. It has been shown in various cases that the obtained series is generally convergent~\cite{Atkinson60,Kay61}, %A ref added
and that the convergence is usually quite fast~\cite{Berry72,Berry82}. In Fig.~\ref{GraphO1_Fig}, we give a diagrammatic representation of the perturbative series. In a regime where the various scattering coefficients are small, $|\alpha| \ll 1$, $|\beta| \ll 1$, and $|R| \ll 1$, it is legitimate to truncate this series at first order, since higher orders will be essentially given by higher products in these quantities. This is true in the smooth limit $\grad \ll 1$, but we also need $\om$ to be sufficiently far from $\omin$, otherwise we would have $|\alpha| = O(1)$ (although $\omin - \om$ might in practice be quite small and $|\alpha| \ll 1$ still valid, see e.g. Fig.~\ref{AlphaBeta_Fig}). We now assume that this is the case, and study the consequence of the first order result \eqref{AlphaInt}. 

The main contribution of the integral governing \eq{AlphaInt} comes from the \emph{saddle point} of the exponential. This saddle point satisfies the equation 
\be \label{Eq_sp}
k_\s(\xctp) - k_\+(\xctp) = 0. 
\ee
By assumption, this equation is not verified by any real $x$. However, when the background functions are analytic, there exist \emph{complex solutions} $x_* \in \mathbb C$. If $x_*$ were real, it would correspond to a turning point, and, hence, in our case we call $x_*$ a \emph{complex turning point}. Decomposing it in real and imaginary parts,  
\be
\xctp = \xr^\alpha + i \Delta^\alpha. 
\ee
It follows from \eqref{AlphaInt} that the $\alpha$ coefficient is given by the contribution of the saddle point as  
\be \label{Alpha}
\alpha \sim C \exp\left({i \int_{x_0}^{\xr^\alpha + i \Delta^\alpha} \left(k_\s(x') - k_\+(x')\right) dx'}\right), 
\ee
where $C$ is a constant prefactor (discussed below). In \eq{Alpha}, $x_0$ is a real reference point, which can be chosen anywhere. If several turning points are present, $\alpha$ is given by a sum of the contribution \eqref{Alpha} for each of them. Usually, the ones that are the closest to the real axis give the dominant contributions, while the others produce only exponentially small corrections~\footnote{\label{Undulation_ftn} In the presence of an undulation in the background, we believe that a series of turning points, corresponding to the bumps of the undulation, will contribute to the scattering coefficients. Hence, these should be radically reduced or increased depending on whether the transition ``resonates'' with the undulation. We believe that this point deserves further studies.}. As we shall see in Sec.~\ref{GenProf2_Sec}, very symmetric flows typically have two main interfering contributions to the scattering coefficients. Taking the modulus of \eqref{Alpha}, we have 
\be \label{Alphamod}
|\alpha|^2 \sim |C|^2 \exp\left(-2 \Im\left( \int^{\xr^\alpha + i \Delta^\alpha}_{x_0} \left(k_\s(x') - k_\+(x')\right) dx' \right) \right). 
\ee
As explained in App.~\ref{prefact_App}, the complex turning point must be chosen such that the contour integral in \eqref{Alphamod} has a positive imaginary part. It follows that $\alpha$ is generally exponentially small, which is a common feature of low gradients or adiabatic limits~\cite{Berry72,Davis76}. The prefactor $C$ in \eqref{Alphamod} is rather delicate to obtain. In the smooth limit $\grad \to 0$, we show that it tends to 1 (see App.~\ref{prefact_App}). However, this limit fails at reproducing the ultra low frequency behavior of the coefficients. The reason is that the limits $\grad \to 0$ and $\om \to 0$ do not commute. Indeed, when the gradients are nonzero but small, in the limit $\om \to 0$, the prefactor vanishes as 
\be \label{prefact0}
|C|^2 \sim \frac{\om}{\oms}, 
\ee
as we show in App.~\ref{oms_App}. The characteristic frequency $\oms$ is estimated in \eq{oms}. The key point is that $\oms$ is proportional to $\grad$, and, hence, becomes very small in the smooth limit, and in particular, $\oms \ll \omin$ ($\omin$ defined in \eq{omin_eq}). To summarize, the prefactor $C$ is characterized by two regimes. When $\oms \ll \om \lesssim \omin$ it is 1, but for $\om \ll \oms$ it is given by \eq{prefact0}. A similar computation for $\beta$ leads to a similar expression, 
\be \label{Betamod}
|\beta|^2 \sim |C|^2 \exp\left(-2 \Im\left( \int^{\xr^\beta + i \Delta^\beta}_{x_0} \left(k_\s(x') - k_\-(x')\right) dx' \right) \right). 
\ee
The complex turning point for $\beta$ is a priori \emph{different} from that of $\alpha$, since it obeys a different resonance condition $k_\s(\xctp^\beta) - k_\-(\xctp^\beta) = 0$. In App.~\ref{oms_App} we show that when $\oms \ll \omin$, the prefactor is essentially the same as for $\alpha$. As we see from \eqref{Alphamod} and \eqref{Betamod}, in the smooth limit, the scattering coefficients are exponentially small. 

The last coefficient $R$, giving the mode mixing between the two long wavelength modes (see Fig.~\ref{Scat_Fig}), can also be evaluated perturbatively, and possesses a contour integral expression as \eqref{Alphamod} and \eqref{Betamod} involving a different complex turning point $\xctp^R$. The first order expression of $R$ is however more delicate. Indeed, for very low frequencies, this first order expression becomes of order 1, meaning thatx the perturbative treatment breaks down. On the contrary, the expressions for $\alpha$ and $\beta$ stay small in the limit $\om \to 0$. The reason for this discrepancy can be seen in the expressions of the matrix elements of $\mathcal M$. While $\mathcal M_{\co \s}$ is proportional to the background gradients $\grad$, $\mathcal M_{\+ \s}$ and $\mathcal M_{\- \s}$ are further suppressed by a factor $O(\om^{1/2})$. Heuristically, we explain this by the fact that $R$ governs a transition involving only long wavelength modes, while $\alpha$ and $\beta$ involve a short wavelength one, which improves the accuracy of the perturbative treatment even at low frequencies $\om \ll \omin$. Fortunately, it has been numerically obtained in various works~\cite{Macher09b,Michel14,Michel15} that the reflection coefficient $R$ stays small for all frequencies. Since we are mainly interested in the production of short wavelength modes by $k_\s$, we shall ignore the mode $k_\co$ in the sequel. Note that to obtain second order estimates of $\alpha$ and $\beta$, $k_\co$ can no longer be ignored since it will appear as an intermediate state in the transitions $k_\s \to k_\+$ or $k_\s \to k_\-$ (see Fig.~\ref{GraphO1_Fig}).

\section{Application to near critical flows}
\subsection{The simplest example: Case of a short obstacle}
\label{ShortObst_Sec}
We shall start by analyzing a simple example. This will allow us to present the techniques to explicitly evaluate \eq{Alphamod}, in a case where the computations stay relatively simple. For this we assume that the mode mixing essentially takes place in a close vicinity of the top of the obstacle, i.e., where $\Fm$ is reached. In Sec.~\ref{GenProf2_Sec} we will give a more precise meaning to this ``short obstacle limit''. Under this assumption, the evolution of the Froude number is well approximated by a second order Taylor expansion near its maximum 
\be \label{Top_obst}
1-F(x) \simeq 1 - \Fm + \frac12 (x/\Lvar)^2. 
\ee
The parameter $\Lvar$ characterizes the length of variation of the Froude number near its maximum value. As we shall see, this quantity directly affects the scattering coefficient. In addition, to simplify the discussion, we present the results in two steps depending on the ratio $\om/\omin$ (but without assuming anything concerning the ratio $\om/\oms$). We first study the limit $\om/\omin \ll 1$, and in a second part, study the corrections in $\om/\omin$.

\subsubsection{Low frequency limit}
\label{ShortObst0_Sec}
In the limit $\om \ll \omin$, the resonance conditions for $\alpha$ and $\beta$ become the same, and reduce to  
\be \label{Chor}
1 - F(\xctp) = 0. 
\ee
Since this condition gives the location of the horizon when the flow is transcritical, for low frequencies, the complex turning point can be interpreted as a \emph{complex horizon}. Using the profile of \eq{Top_obst}, it is given by  
\be
\xctp = \pm i \Lvar \sqrt{2(1 - \Fm)} = \pm i \Delta_0. 
\ee
To obtain the correct sign of the integral in \eq{Alphamod}, we must choose $\Im(\xctp) > 0$. Moreover, when $\om \to 0$, the roots of the Hamilton-Jacobi equation \eqref{HJ} become simpler, and we find   
\be
k_\s(x) - k_\+(x) = \frac1{\h(x)} \sqrt{3(1 - F^2(x))}. \label{kZ_al} 
\ee
We notice here that the root difference \eqref{kZ_al} scales like $(1-F)^{1/2}$. Therefore, in the limit $1-\Fm \ll 1$, it is only necessary to consider the variations of the function $1-F(x)$. The other background quantities can be approximated by their value near $\Fm$, since taking into account extra terms will produce subleading corrections in $1-\Fm$. In this limit, the profile of \eq{Top_obst} gives  
\be
k_\s(x) - k_\+(x) = \frac1{\Lvar h_\m} \sqrt{3(\Delta_0^2 + x^2)}. 
\ee
We now use this expression to compute the complex integral governing the scattering coefficient $\alpha$ through \eq{Alphamod}. For convenience, we chose the reference point $x_0 = 0$, then  
\bsub \bea
\int^{i \Delta_0}_{0} \left(k_\s(x') - k_\+(x')\right) dx' &=& \frac1{\Lvar h_\m} \int^{i \Delta_0}_{0} \sqrt{3(\Delta_0^2 + x'^2)} dx' , \\ 
&=& i\frac{\sqrt{3} \Delta_0^2}{\Lvar h_\m} \int^{1}_{0} \sqrt{1 - t^2} dt , \\ 
&=& i\frac{\sqrt{3} \pi \Delta_0^2}{4 \Lvar h_\m} . 
\eea \esub
We deduce the amplitude of the $\alpha$ coefficient 
\be \label{Alpha_om0}
|\alpha_0|^2 = |C_\om|^2 \exp\left(-\frac{\sqrt{3} \pi \Lvar}{h_\m} (1-\Fm) \right), 
\ee
where $\alpha_0$ is short for $\alpha_{\om \ll \omin}$. We see that in the regime $\om \ll \omin$, all the frequency dependence is in the prefactor $C_\om$ (we added the index $\om$ with respect to \eq{Alphamod} to emphasize this point). Since we assume nothing concerning the ratio $\om/\oms$, $C_\om$ varies from 1 to $\om/\oms$ when $\om$ decreases. By a similar computation, we show that $\beta_0$ has the same amplitude for very low frequencies, i.e. $|\beta_{\om \to 0}|^2 \sim |\alpha_{\om \to 0}|^2$. This can be seen by direct computation, but comes in fact from a more general property of the mode equation. Indeed, the change $\phi_\om \to (\phi_{-\om})^*$ leaves the mode equation \eqref{EoM} invariant, and as can be seen by looking at the roots of \eqref{HJ}, exchanges the role of $\alpha$ and $\beta$. This leads to the relation 
\be \label{AlBetSym}
\beta_{\om} = \alpha_{-\om}^*. 
\ee
The property above has been widely used in Hawking radiation studies~\cite{Brout95,Balbinot06,Coutant11}. Here also, it significantly simplifies the computations of the scattering coefficients.

\subsubsection{Frequency dependence}
\label{OmDep_Sec}
\begin{figure}[!ht]
%\begin{center}
\subfigure[]{\includegraphics{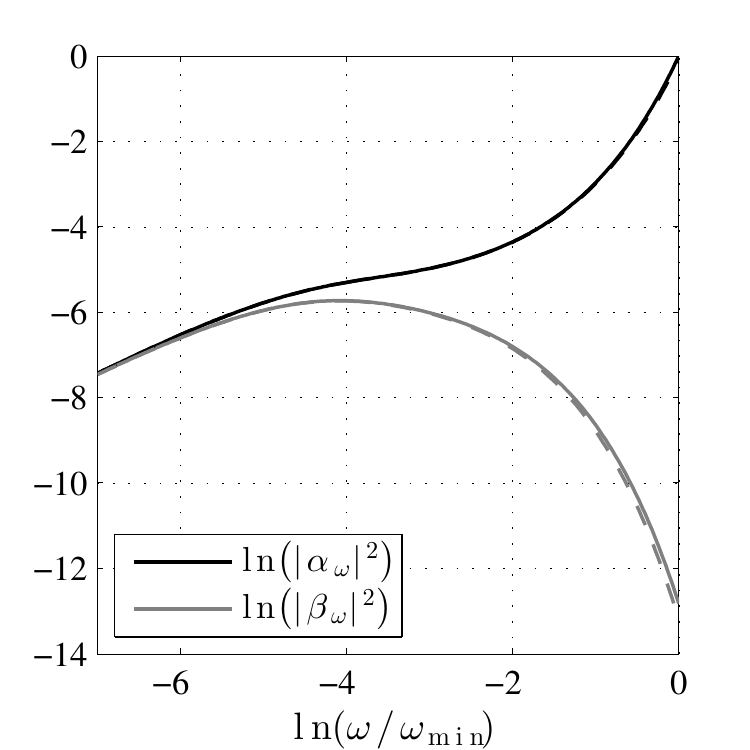}\label{AlphaBeta_Fig}}
\subfigure[]{\includegraphics{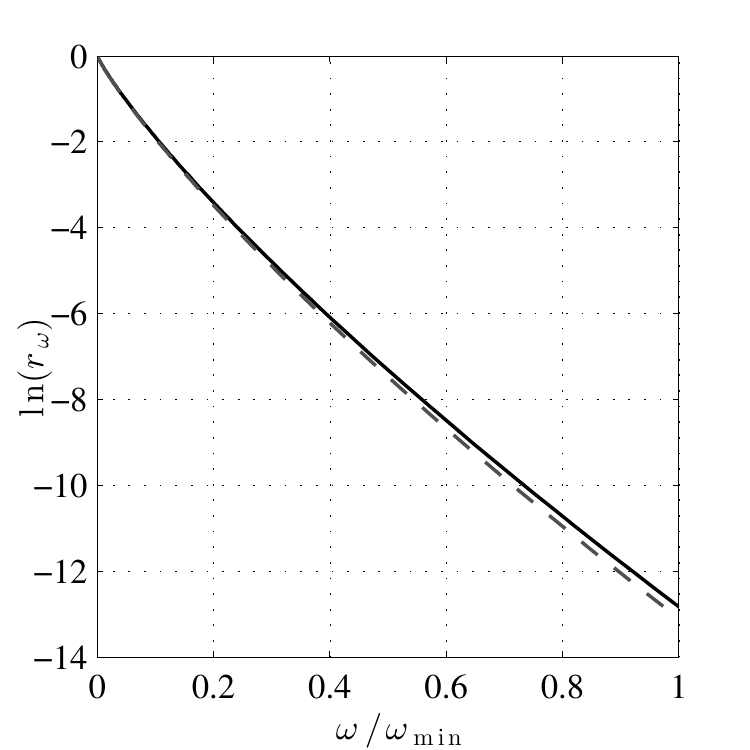}\label{Ratio_Fig}}
%\end{center}
\caption{Left panel (a): coefficients $\alpha_\om$ (black) and $\beta_\om$ (grey) as a function of $\om$, as given by \eq{Alpha_om}. The prefactor $C_\om$ is given by \eq{Ceff}. We clearly observe three distinct regimes: for $\om \ll \oms$, $|\alpha_\om|^2\sim |\beta_\om|^2 \propto \om$, for $\oms \ll \om \ll \omin$, $|\alpha_\om|^2\sim |\beta_\om|^2$ almost constant in $\om$, and for $\oms \ll \om \lesssim \omin$, $|\alpha_\om|^2$ increases, while $|\beta_\om|^2$ decreases. Right panel (b): Ratio $r_\om$ as a function of $\om$. 
In both plots, we have chosen the flow parameters such that $\Lvar (1-\Fm)/h_\m =1 $, and $\ln(\oms/\omin) \simeq -4.4$. The solid lines are obtained using the full functions $\mathcal I_\pm$, while the dashed ones are the approximations \eqref{AlphaBetaLowom}, and \eqref{RatioLowom}. Note that the present treatment cannot be trusted too close to $\ln(\om/\omin) \approx 0$. 
}
\end{figure}

The corrections to \eq{Alpha_om0} in $\om/\omin$ are more delicate to obtain. These corrections have two origins. The first is the shift of the value of the complex turning point, and the second is the exact expression of the roots $k_\s$, $k_\+$, and $k_\-$. For the latter, one needs to solve the Hamilton-Jacobi equation \eqref{HJ}. Unfortunately, one cannot simply compute the corrections to \eq{kZ_al} for $\om \ll \omin$ because such corrections will not be accurate close to the turning point.  We can still simplify \eq{HJ} by discarding the last root $k_\co$, which plays essentially no role at first order in perturbation theory. The Hamilton-Jacobi equation \eqref{HJ} is then reduced to a third order equation in $k$. To obtain it, we carefully take the square root of \eqref{HJ} so as to select the relevant branch (see Fig.~\ref{HJ_Fig}), and expand the result up to $O(k^3)$. This gives 
\be \label{HJ3rd}
\om = -c(1-F) k + \frac{c\h^2}{6} k^3. 
\ee
A direct comparison of this equation with \eq{HJ} shows that the three roots $k_\s$, $k_\+$, and $k_\-$ are approximated by the roots of \eqref{HJ3rd} up to small corrections in $1-F \ll 1$~\footnote{To see this, we first notice that since we performed an expansion in $k$, the highest error made is on the value of $k_{-}(\om=0)$, which has the highest value (see Fig.~\ref{HJ_Fig}). From \eq{HJ}, it is given by $\h^{-1} \sqrt{3(1-F^2)}$, while \eq{HJ3rd} gives $\h^{-1} \sqrt{6(1-F)}$, which agree whenever $1-F \ll 1$. It is also noticeable that \eq{HJ3rd} corresponds to the dispersion relation of the linearized Korteweg-de Vries equation~\cite{Johnson}.}. We now obtain the roots by solving this equation using the Cardan-Tartaglia method. To start, the associated discriminant gives the equation for all the complex turning points, i.e. it gives the condition for two roots to merge, 
\be \label{tp_Eq}
(1-F(x_*))^3 = \frac{9 \om^2 \h^2(x_*)}{8c^2(x_*)}. 
\ee
Similarly to \eq{kZ_al}, at leading order in $1-\Fm$, it is only necessary to consider the variations of $1-F(x)$, while $\h$ and $c$ are well approximated by $c_\m$ and $h_\m$. 
Doing so, the complex turning point for $\alpha$ is given by 
\be \label{Alpha_tp}
\xctp^\alpha = i \Delta_0\left(1 - \left(\frac{\om}{\omin}\right)^{2/3}\right)^{1/2} . 
\ee
The other complex turning point $\xctp^\beta$ is another solution of \eq{tp_Eq}. For $\om \neq 0$, both emerge from the complex horizon $\xctp^0$, the difference scaling like $O((\om/\omin)^{2/3})$. To simply express the roots, we introduce the auxiliary functions 
\be
U_\om^\pm(x) = \left({\frac1{\h^3} \sqrt{8(1-F)^3 - \frac{9 \om^2 \h^2}{c^2}} \pm i \frac{3 \om}{c \h^2}} \right)^{1/3}.
\ee
The Cardan-Tartaglia method then gives the three roots as combinations of $U_\om^+$ and $U_\om^-$. In particular,  
\be \label{Tart_s+}
k_\s(x) - k_\+(x) = \sqrt{3} e^{i \frac{\pi}3} U_\om^+(x) + \sqrt{3} e^{-i \frac{\pi}3} U_\om^-(x) . 
\ee
We now evaluate this near the top of the obstacle, that is, using \eq{Top_obst}. At leading order in $1-\Fm$, we have 
\be
U_\om^\pm(x) = \frac{\Delta_0}{\Lvar h_\m} \left[\sqrt{\left(1 + \frac{x^2}{\Delta_0^2} \right)^{3} - \left(\frac{\om}{\omin}\right)^{2}} \pm i \frac{\om}{\omin} \right]^{1/3} .
\ee
We are now ready to compute the complex integral governing the coefficient $\alpha$ in \eq{Alphamod}. We start by writing   
\be \label{Uint}
\int_0^{\xctp^\alpha} U_\om^\pm(x') dx' = -i \frac{\Delta_0^2}{\Lvar h_\m} \mathcal I_\pm \left(\frac{\om}{\omin}\right), 
\ee
where we defined the functions $\mathcal I_\pm$ by 
\be
\mathcal I_\pm(\eps) = \int_0^{\sqrt{1-\eps^{2/3}}} \left( \sqrt{\left(1 - t^2 \right)^{3} - \eps^2} \pm i \eps\right)^{1/3} \, dt . 
\ee
Combining the preceding results, and applying \eq{Alpha}, we finally obtain 
\be \label{Alpha_om}
\alpha_\om = C_\om \exp\left(- \frac{2 \sqrt{3} \Lvar (1-\Fm)}{h_\m} \left(e^{i \frac{\pi}3} \mathcal I_+ \left(\frac{\om}{\omin}\right) + e^{-i \frac{\pi}3} \mathcal I_- \left(\frac{\om}{\omin}\right)\right)\right).
\ee
This gives the expression of $\alpha_\om$ in the flow profile of \eq{Top_obst}. \eq{Alpha_om} should be valid up to $\om \lesssim \omin$, as long as $|\alpha_\om| \ll 1$. By using the same method, \eq{Betamod} leads to a similar expression for the coefficient $\beta_\om$. To obtain it, one can either redo the calculation of the complex integral, or more quickly, carefully apply \eq{AlBetSym}. The results are presented on Fig.~\ref{AlphaBeta_Fig}. Since the full expression of the $\mathcal I_\pm$ functions is rather complicated, it is instructive to look at the limit $\om \to 0$, and see how $\alpha_\om$ (resp. $\beta_\om$) deviates from \eq{Alpha_om0}. Interestingly, the functions $\mathcal I_\pm$ are not differentiable for $\eps \to 0$ and therefore, small frequency corrections display non-analytic terms. Indeed, after some efforts, one can show that 
\be
\mathcal I_+(\eps) = \frac{\pi}4 - \frac{\eps}3 - \frac{i\eps}{9} \ln\left(\frac{i \eps}{24}\right) + o(\eps). 
\ee
This gives approximate expressions for the scattering coefficients 
\bsub \label{AlphaBetaLowom} \bea
\ln(|\alpha_\om|^2) &\sim& \ln(|C_\om|^2) - \frac{\sqrt{3} \pi \Lvar (1-\Fm)}{h_\m} \left( 1 - \frac{12 - 2\pi - 4\sqrt{3} \ln(\om/24 \omin)}{9 \pi}  \frac{\om}{\omin} \right) , \\
\ln(|\beta_\om|^2) &\sim& \ln(|C_\om|^2) - \frac{\sqrt{3} \pi \Lvar (1-\Fm)}{h_\m} \left( 1 + \frac{12 + 2\pi - 4\sqrt{3} \ln(\om/24 \omin)}{9 \pi} \frac{\om}{\omin} \right) . 
\eea \esub
As we observe on Fig.~\ref{AlphaBeta_Fig}, the low frequency expressions \eq{AlphaBetaLowom} are quite accurate up to $\om \lesssim \omin$ (where the perturbative expression \eqref{Alphamod} can no longer be trusted). At this level we would like to emphasize several qualitative features displayed by \eq{AlphaBetaLowom} that are maintained for more general profiles. First, when $\om \to 0$, $|\alpha_\om|^2 \sim |\beta_\om|^2$ and both vanish as $O(\om)$ due to the prefactor (see \eq{prefact0}). Second, when $\om/\omin$ increases, $|\alpha_\om|^2$ becomes larger than $|\beta_\om|^2$. Third, the corrections in $\om/\omin$ display non-analytic terms, in $O(\om \ln(\om))$. 

It is also quite instructive to analyse the behavior of the ratio $r_\om = |\beta_\om/\alpha_\om|^2$. Indeed, the linearity of the logarithm of this ratio in $\om$ has been used in the literature as a sign for the thermality of the emitted spectrum. Moreover, this ratio is also independent of the prefactor $C_\om$. For low frequencies, \eq{AlphaBetaLowom} gives 
\be \label{RatioLowom}
\ln(r_\om) \sim - \frac{\sqrt{3} \Lvar (1-\Fm)}{h_\m} \left( \frac{24 - 8\sqrt{3} \ln(\om/24 \omin)}{9}  (\om/\omin) \right). 
\ee
On Fig.~\ref{Ratio_Fig}, we plotted the evolution of $r_\om$, using both \eq{Alpha_om} and the low frequency expression \eqref{RatioLowom}. As we see, despite the presence of non-analytic corrections, $r_\om$ looks fairly linear in $\om$. However, this ratio alone misses several features of the scattering that differs from the Hawking regime, and in particular the low frequency $\om \ll \omin$ behavior of $\alpha$ and $\beta$. 

\subsection{A general class of flow profiles}
\subsubsection{Monotonic profiles}
We shall start by analyzing the case of a profile whose Froude number, increase monotonically from a minimum to a maximum value. This case is very useful to better understand the more general profiles of Sec.~\ref{GenProf2_Sec}, but it also has its interests in its own right. To model such a profile, we assume that the Froude number is given by 
\be \label{1StepProf}
F(x) = F_0 + D \tanh\left(\frac{\gamma x}D \right). 
\ee
The flow starts from a low Froude number $F_\m = F_0 - D$ on the left side, and smoothly rises to reach $\Fm = F_0 + D$. The parameter $\gamma$ gives the slope of the profile. As in the preceding section, it is simpler to first look at the low frequency limit $\om \ll \omin$ and in a second time discuss the corrections in $\om/\omin$. At low frequencies, the physics is dictated by the complex horizon, i.e. the location satisfying \eq{Chor}, which governs the common value of $\alpha_0$ and $\beta_0$. From equation \eqref{1StepProf}, we find the complex horizon~\footnote{There is in fact a discrete periodic set, with imaginary parts that are odd multiples of the one of \eqref{1StepHor}. We keep here only the one closest to the real axis, which gives the dominant contribution. There are also poles located at $ i (2n+1)\pi D/2\gamma$, but a direct (similar to App.~\ref{Contour_App}) calculation of the corresponding contributions shows that they are subdominant with respect to the complex horizon \eqref{1StepHor}.} to be 
\be \label{1StepHor}
\xctp^0 = \frac{D}{2\gamma} \ln\left(\frac{1-F_\m}{1-\Fm}\right) + i \frac{\pi D}{2\gamma} . 
\ee
We then compute the low frequency value of $\alpha$ and $\beta$ (see App.~\ref{Contour_App}), and we find 
\be \label{1StepAlpha}
|\alpha_0|^2 \sim |\beta_0|^2 \sim |C_\om|^2 \exp\left(- \frac{\sqrt{6} \pi D}{\gamma h_\m} (1-\Fm)^{1/2}\right). 
\ee
We see that the value of the coefficient depends not only on the slope $\gamma$, but also on the height of the step, i.e. the parameter $D$. Moreover, we notice that it depends on $1-\Fm$ with a different power law than in the short obstacle case (compare \eqref{1StepAlpha} to \eqref{Alpha_om0}). The reason for this is that unlike in the short obstacle case, the imaginary part of the complex horizon of \eqref{1StepHor} is independent of $1-\Fm$, hence $\alpha_0$ depends on it only through the roots $k_\s - k_\+$. 
To obtain the corrections for $\om \neq 0$, we follow the same procedure as in Sec.~\ref{OmDep_Sec}, and compute only the leading order correction. A rather tedious computation shows that for small $\om/\omin$, and $1-\Fm \ll 1$,  
\bsub \label{1StepAlphaom} \bea
\ln(|\alpha_\om|^2) &=& \ln(|C_\om|^2) - \frac{\sqrt{6} \pi D (1-\Fm)^{1/2}}{\gamma h_\m} \left(1 - \frac{\sqrt{3} \om}{3 \omin}\right) , \\
\ln(|\beta_\om|^2) &=& \ln(|C_\om|^2) - \frac{\sqrt{6} \pi D (1-\Fm)^{1/2}}{\gamma h_\m} \left(1 + \frac{\sqrt{3} \om}{3 \omin}\right) .
\eea \esub
We notice that unlike in \eqref{AlphaBetaLowom}, the above equation shows no non-analytic terms. This turns out to be an accident of the profile of \eq{1StepProf}, where the leading order non-analytic terms (in $O(\om \ln(\om))$) contributes only to the phase of $\alpha_\om$. This is no longer true for the next-to-leading corrections in $\om/\omin$ or if the profile slightly differs from \eqref{1StepProf}.

\subsubsection{Non-monotonic profiles}
\label{GenProf2_Sec}
\begin{figure}[!ht]
\includegraphics{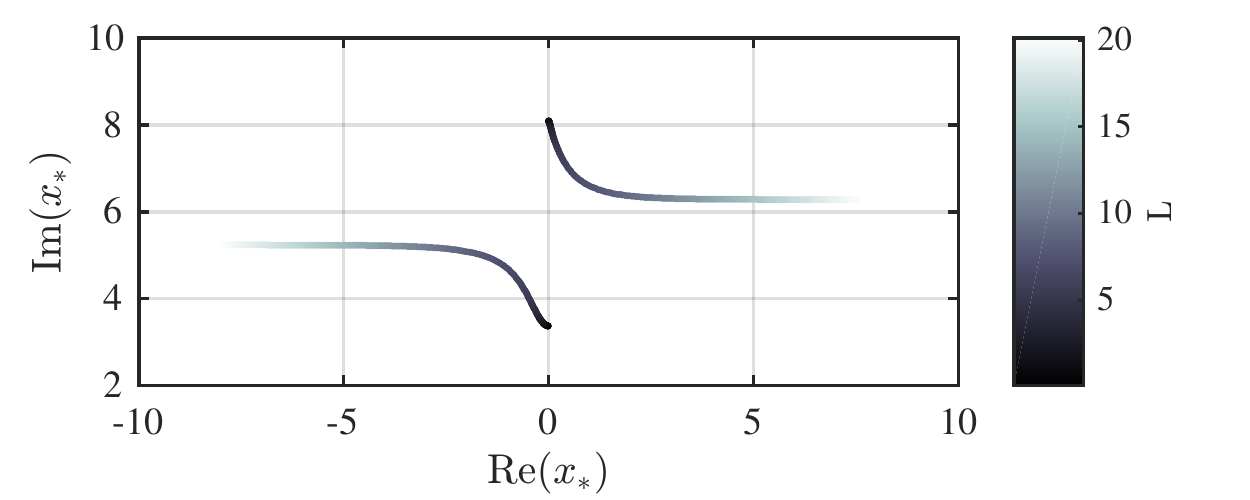}
\caption{Evolution of the two complex horizons solutions of \eqref{Chor} for the profile \eqref{GeneralProf} as $L$ varies. The profile is slightly asymmetric: $\gamma_l = 0.3 D$ and $\gamma_r = 0.25 D$, hence in the long obstacle limit, i.e. large $L$, one of them lies closer to the real axis. The other parameters of the flow \eqref{GeneralProf} are $F_0 = 0.71$, and $D=0.17$. 
}
\label{Chor_Fig} 
\end{figure}

We are now ready to analyze a more general class of flow profile, which have a similar shape than the ones studied numerically and experimentally~\cite{Weinfurtner10,Michel14,Michel15,Euve15} (undulation excluded). For this we assume that the Froude number is given by 
\be \label{GeneralProf}
F(x) = F_0 - D \tanh\left(\frac{\gamma_l (x+ L/2)}D \right) \tanh\left(\frac{\gamma_r (x-L/2)}D \right) . 
\ee
The maximum value of the Froude number $\Fm$ is by assumption smaller than 1. The flow starts from a low Froude number $F_\m = F_0 - D$ on the left side, rises to reach $\Fm$ and then decreases again to $F_\m$. On the left side (resp. right side), the slope is controlled by the parameter $\gamma_l$ (resp. $\gamma_r$). 
Again, we first consider the low frequency limit $\om \ll \omin$ and then discuss the corrections in $\om/\omin \neq 0$. Depending on the parameters of the profile, we distinguish three different regimes:
\bi
\item long obstacles $\gamma_{l, r} L \gg 1$, asymmetric $\gamma_l \neq \gamma_r$, 
\item long obstacles $\gamma_{l, r} L \gg 1$, symmetric $\gamma_l = \gamma_r$, 
\item short obstacles $\gamma_{l, r} L \ll 1$. 
\ei
When the obstacle is long ($L$ is large), there exist \emph{two} complex horizons, located around the top of each slopes. In the limit $\gamma_{l, r} L \gg 1$, they are given by the monotonic profile expression \eqref{1StepHor} centered at $\pm L/2$, i.e. 
\bsub \label{doubleChor} \bea
\xctp^l &=& -\frac L2 + \frac{D}{2\gamma_l} \ln\left(\frac{1-F_\m}{1-\Fm}\right) + i \frac{\pi D}{2\gamma_l} , \\
\xctp^r &=& \frac L2 - \frac{D}{2\gamma_r} \ln\left(\frac{1-F_\m}{1-\Fm}\right) + i \frac{\pi D}{2\gamma_r} . 
\eea \esub
The scattering coefficients are then given by a sum of two interfering contributions 
\be \label{interferedAlpha}
|\alpha_\om|^2 \sim |\alpha^l|^2 + |\alpha^r|^2 + 2|\alpha^l||\alpha^r| \cos\left(\Re\int_{\xctp^l}^{\xctp^r} (k_\s(x') - k_\+(x')) dx' \right), 
\ee
where $\alpha^{l, r}$ are given by the single turning point expression \eqref{1StepAlpha} with $\gamma = \gamma_{l, r}$. This equation allows us to draw several conclusions concerning the behavior of the scattering coefficient. If the profile is asymmetric, $\gamma_l \neq \gamma_r$, the one with the \emph{biggest} slope dominates in the expression for $|\alpha_\om|^2$. Indeed, the imaginary part of the corresponding turning point lies closer to the real axis, as it is inversely proportional to $\gamma_{l, r}$. On the contrary, if the flow is symmetric, $\gamma_l = \gamma_r$, the two contributions have the same weight and interfere through the phase of \eq{interferedAlpha}. This phase shift is accumulated not only along the real line, between $\xr^l$ and $\xr^r$, but also in the complex plane, from $\xr^{l, r}$ to $\xctp^{l, r}$. For a long obstacle ($\gamma_{l, r} L \gg 1$) and $\om \ll \omin$, the phase shift of \eqref{interferedAlpha} is given by 
\be \label{ResPhase}
\Re\int_{\xctp^l}^{\xctp^r} (k_\s(x') - k_\+(x')) dx' = \zeta_l - \zeta_r + \int_{\xr^l}^{\xr^r} \frac1{\h} \sqrt{6(1-F(x'))} dx', 
\ee
where $\zeta_{l,r}$ are defined after \eq{result_contour}. When decreasing $L$, the two complex horizons keep the same imaginary part, but their real part get closer. At a certain critical value $L = L_{\rm c}$, they merge into a single solution~\footnote{Using \eq{GeneralProf}, the critical value for $\gamma_l = \gamma_r$ can be shown to be $L_{\rm c} = 2 D \gamma^{-1} \mathrm{artanh}(\sqrt{D/(1-F_0)})$.}. For lower values $L < L_{\rm c}$, one of the solutions migrates closer to the real axis, while the other moves afar. If the profile is not perfectly symmetric, one observes something similar, but instead of merging together, the roots first get closer, and around the critical value of $L$, repel each other so that the one with the smallest imaginary part approaches the real axis, while the other moves afar (see Fig.~\ref{Chor_Fig}). This mechanism is very similar to the ``\emph{avoided crossing}'', well-known in quantum mechanics~\cite{Gottfried}. When $L < L_{\rm c}$, we enter in the regime of a short obstacle. In this case, one of the complex horizons dominates in the expression \eqref{interferedAlpha} for $\alpha$. This solution has a real part close (equal if $\gamma_l = \gamma_r$) to zero, i.e., it lies close to the top of the obstacle. This case becomes very similar to the one studied in Sec.~\ref{ShortObst_Sec}. At $L = 0$, the complex horizon closest to the real axis is found to be 
\be \label{tp_ch}
\xctp = \frac{i D}{\gamma} \arcsin \left(\sqrt{\frac{1-\Fm}{1-F_\m}}\right) . 
\ee
If we additionally have $1-\Fm \ll 1-F_\m$, \eqref{tp_ch} exactly reduces to the short obstacle case of Sec.~\ref{ShortObst_Sec}, meaning that \eq{Top_obst} becomes a good approximation to describe the scattering. 

\begin{figure}[!ht]
\subfigure[]{\includegraphics{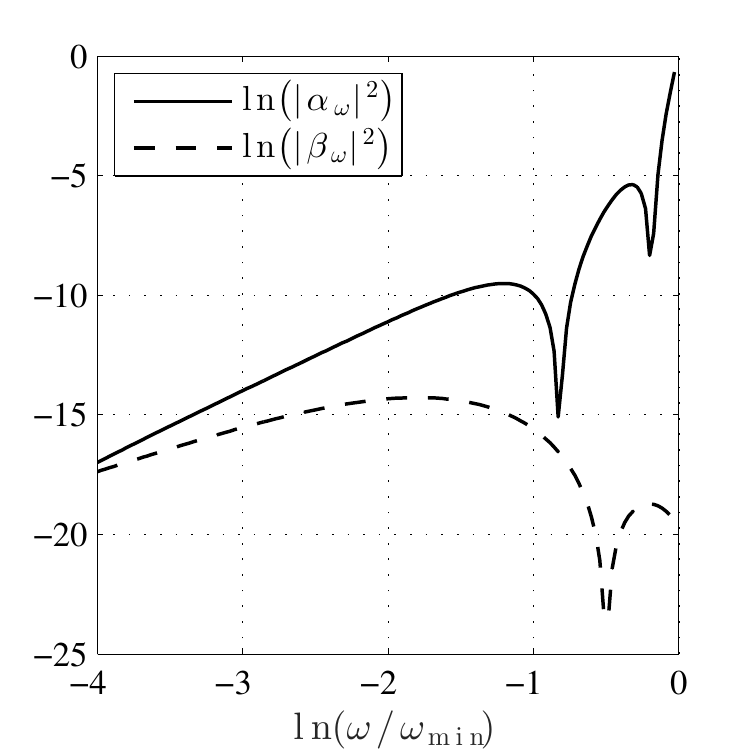}}
\subfigure[]{\includegraphics{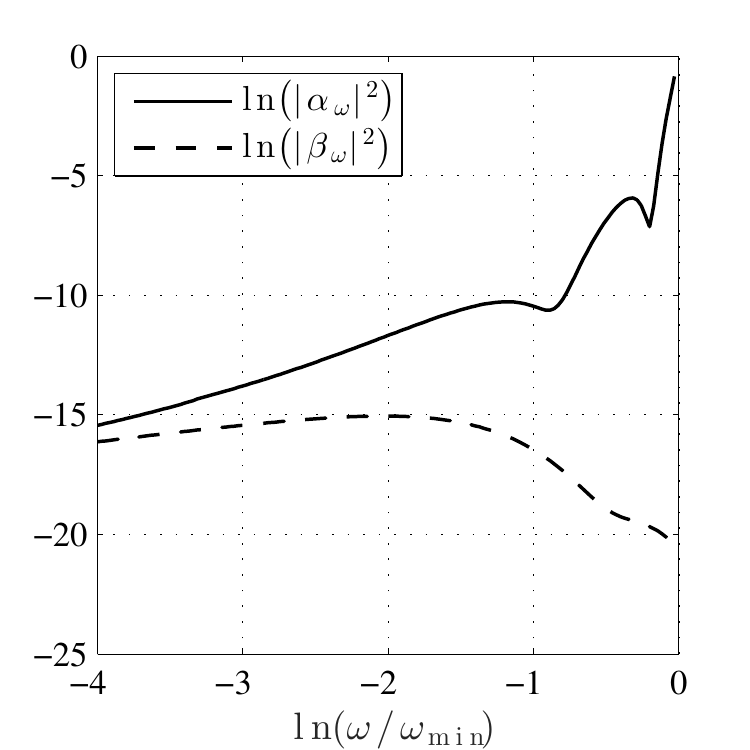}}
\caption{Coefficients $\alpha_\om$ (solid) and $\beta_\om$ (dot-dashed) as a function of $\om$ for a long obstacle $\gamma_{l, r} L \gg 1$. As in Eqs.~\eqref{AlphaBetaLowom} and \eqref{1StepAlphaom}, we have taken into account the leading corrections in $\om/\omin$. We work in units where $g=q=1$ and assume that \eq{IdealFlow} holds for simplicity. The parameters of the flow \eqref{GeneralProf} are $F_0 = 0.7$, $D=0.17$ and $L=25$. With these parameters, $\ln(\oms/\omin) \simeq -1.8$. The prefactor $C_\om$ is given by \eq{Ceff}. Left panel (a): symmetric profile, $\gamma_l = \gamma_r = 0.2 D$. Right panel (b): asymmetric profile, $\gamma_l = 0.2 D$ and $\gamma_r = 0.17 D$. As we see, a small asymmetry quickly reduces the interference effects. 
}
\label{2StepAlphaBeta_Fig} 
\end{figure}

\begin{figure}[!ht]
\subfigure[]{\includegraphics{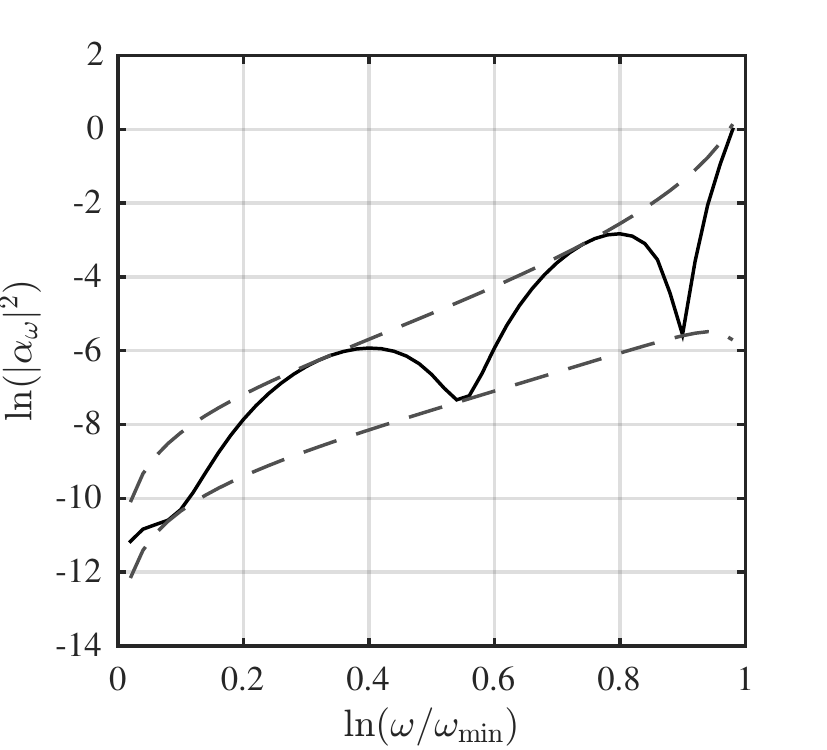}} 
\subfigure[]{\includegraphics{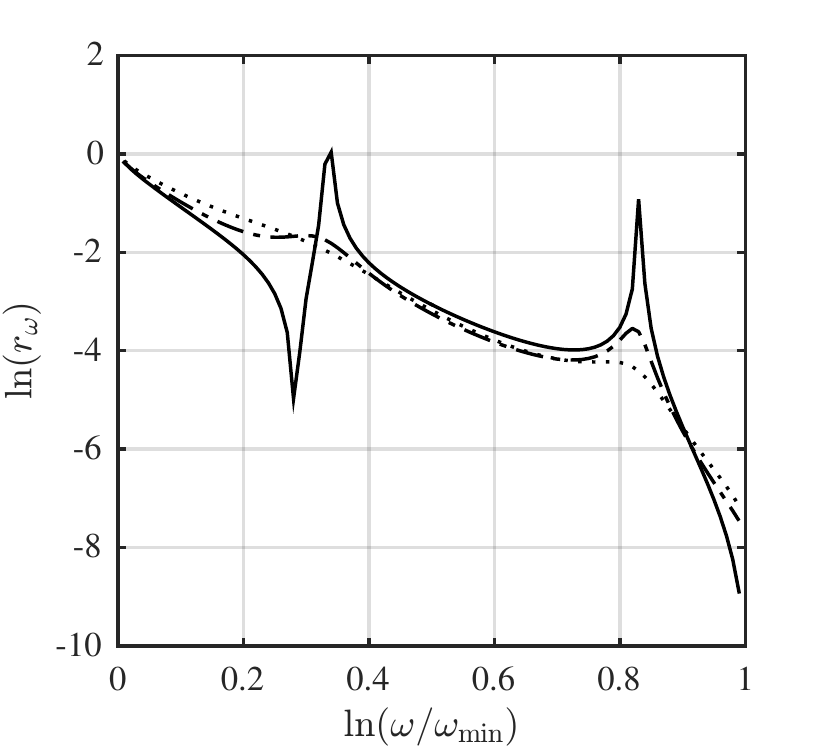}} 
\caption{Left panel (a): coefficient $\alpha_\om$ (solid) as a function of $\om$ for a long obstacle. We represented the extreme values $(|\alpha^l|+|\alpha^r|)^2$, and $(|\alpha^l|-|\alpha^r|)^2$ (dashed) to emphasize the effect of interferences. Right panel (b): plot of the ratio $r_\om = |\beta_\om/\alpha_\om|^2$ as a function of $\om$, for three long obstacles less and less symmetric: $\gamma_r/\gamma_l = 1$ (solid), $\gamma_r/\gamma_l = 1.2$ (dot-dashed), $\gamma_r/\gamma_l = 1.5$ (dotted), all with $\gamma_l = 0.3 D$. We see that when the obstacle becomes asymmetric, the ratio becomes fairly linear, as in the case exposed in Fig.~\ref{Ratio_Fig}. The other parameters of the flow \eqref{GeneralProf} are $F_0 = 0.71$, $D=0.17$ and $L=25$.
}
\label{2StepOscill_Fig} 
\end{figure}

When relaxing the assumption $\om \ll \omin$, $\alpha_\om$ and $\beta_\om$ differ but are still given by an interfering sum as in \eq{interferedAlpha}. 
As $\om$ increases, the two turning points $\xctp^\alpha$ and $\xctp^\beta$ emerge from $\xctp^0$ and migrate in different directions in the complex plane. The effect of this migration is twofold. First, for long obstacles, the relative location of the left and right turning point changes, and therefore, the phase \eqref{ResPhase} between the two interfering contributions in \eq{interferedAlpha} varies. For some values of the frequency, this phase will be a multiple of $2\pi$, and the coefficients show a dip, as the first order estimate in \eq{interferedAlpha} vanishes. Such dips have been numerically observed in~\cite{Michel14}. Second, the amplitudes of the single turning point contributions, i.e. $|\alpha^l|$ and $|\alpha^r|$ (resp. $|\beta^l|$ and $|\beta^r|$ for $\beta_\om$) are altered as in \eq{1StepAlphaom}. On Fig.~\ref{2StepAlphaBeta_Fig}, we represented the evaluation of $\alpha_\om$ and $\beta_\om$ for a symmetric and an asymmetric profile. On Fig.~\ref{2StepOscill_Fig}, we show how the coefficients oscillate in $\om$ due to interferences between the two turning points as in \eq{interferedAlpha}, and how this affects the ratio $r_\om$. 

%%%%%%%%%%%%%%%%%%%%%%%%%%%%%%%%%%
% CONCLUSION
%%%%%%%%%%%%%%%%%%%%%%%%%%%%%%%%%%
\section{Conclusion}

In this paper, we studied the scattering of low frequency waves on a subcritical fluid flow, that is, whose Froude number stays below 1. We developed a new method, based on a generalization of the Bremmer series, where exact solutions of the wave equation are written as a local superposition of WKB modes (see \eq{Adiab0}). The coefficients of this superposition, which we called local scattering coefficients, are position dependent and possess several useful properties. First, they are by construction slowly varying. At some locations along the flow, they can transit from one constant value to another. This can be interpreted as the creation of a new mode. Second, their asymptotic values directly give the scattering coefficients. Third, the local scattering coefficients are governed by a first order differential equation, \eq{Scat_Eq}, which is equivalent to the original wave equation and is adapted to a perturbative treatment at low gradients, i.e. $\grad \ll 1$. 

In Sec.~\ref{tp_Sec}, we expose the first order perturbative results of this series. We show that the coefficients are mainly governed by {complex turning points}, corresponding to the locations where two roots of the dispersion relation \eqref{HJ} merge when these are analytically continued in the complex plane. In general, there exist many turning points in the complex plane. Importantly, the ones closest to the real axis dominate while the others contribute as exponentially small corrections. Hence the scattering coefficients are governed by a few dominating contributions, taking the form of complex exponentials of contour integrals from the real line to the complex turning points, see Eqs.~\eqref{Alphamod}, and \eqref{Betamod}.  

We then applied these results to a large class of flow profiles, so as to extract the generic features of the scattering coefficients $\alpha$ and $\beta$. By studying the behavior of the scattering coefficients as a function of the frequency $\om$, we distinguish three main regimes. For ultra-low frequencies, $\om \ll \oms$, $|\alpha|^2$ and $|\beta|^2$ both vanish linearly in $\om$, see \eq{prefact0}. For intermediate frequencies $\oms \ll \om \ll \omin$, $|\alpha|^2$ and $|\beta|^2$ share a constant value as in \eq{Alpha_om0}, and when $\om/\omin$ becomes significant, they start drifting apart as shown by \eq{AlphaBetaLowom}. When they do, $|\beta|^2$ is generically smaller than $|\alpha|^2$. Moreover, we show in Sec.~\ref{GenProf2_Sec} that long obstacles generally produce two dominating complex turning points. If the obstacle is symmetric enough, these two contributions give rise to oscillations in $|\alpha|^2$ and $|\beta|^2$ due to interferences, as illustrated in Figs.~\ref{2StepAlphaBeta_Fig}, and \ref{2StepOscill_Fig}. All these features are in perfect agreement with what have been previously observed numerically in~\cite{Finazzi11,Michel14,Michel15} and are complemented by analytic predictions for the parameters governing the various regimes. 

In all, this analysis describes in detail what is the ``imprint'' of Hawking radiation when the flow accelerates but stays subcritical. The physics of the Hawking effect is dictated by horizons, and we have shown here that its imprint in subcritical flows is governed by \emph{complex turning points}. In this regime, the spectrum becomes more complicated, as it is governed by nonlocal quantities. The study of complex turning points allowed us to provide a simple characterization of this spectrum. When increasing the Froude number, these turning points get closer to the real axis, until they reach it. Before they do, the present treatment breaks down, but it is expected that for increasing $F$ the spectrum will smoothly change from the subcritical one to the Hawking one when $F$ is sufficiently larger than 1. In the Hawking regime, the characteristic length of non-locality becomes smaller than the characteristic length of the gradients~\cite{Coutant14b}, and as a result, the spectrum becomes entirely governed by the surface gravity, i.e. the gradient of the Froude number at the horizon. The analytical study of this transition will be the aim of future investigations. 

\acknowledgements
We would like to thank Florent Michel, Renaud Parentani, Scott Robertson, and Bill Unruh for useful comments about the final version of this manuscript. We also thank Michael Berry for discussions. This project has received funding from the European Union's Horizon 2020 research and innovation programme under the Marie Sk\l odowska-Curie grant agreement No 655524. S.W. acknowledges financial support provided under the Royal Society University Research Fellow (UF120112), the Nottingham Advanced Research Fellow (A2RHS2) and the Royal Society Project (RG130377) grants.

%%%%%%%%%%%%%%%%%%%%%%%%%%%%%%%%%%
% APPENDIX
%%%%%%%%%%%%%%%%%%%%%%%%%%%%%%%%%%
\newpage
\appendix
\section{Generalized Bremmer series}
\label{Math_App}
\subsection{Bremmer series for equations of order $N$}
\label{Bremmer_App}

In this appendix, we derive the equation satisfied by the local scattering coefficients defined in Sec.~\ref{Bremmer_Sec}. The method we present is a generalization of the Bremmer series~\cite{Bremmer51,Landauer51,Winitzki05}. Whereas the Bremmer series deals with second order differential equations, such as the Schrödinger equation, we consider higher order differential equations~\cite{Atkinson60,Kay61,Keller62}. This is essential to describe dispersive effects of wave propagation, as in our case~\footnote{Since we adopt here the point of view of dispersive wave equations, our approach has several technical differences with respect to other generalizations of Bremmer series~\cite{Atkinson60,Kay61,Keller62}, such as the distinction between the $f_n$ and $g_n$, or the adiabatic invariant of $\lam$-canonical systems (see below).}. %A remarks answering Ref C
Higher order differential operators are also useful to study Schrödinger types of equation in momentum representation, see e.g.~\cite{Delos72II,Coutant14,Coutant15}. Our method also bears many similarities with the adiabatic series used in a wide variety of contexts, such as electronic transitions in molecular collisions~\cite{Davis76} or particle creation in cosmology~\cite{Massar97}. However, here the corresponding operator is not self-adjoint. Under certain conditions, there exists a quadratic conserved quantity, but it has no reason to be positive definite. This is the case for the wave equation of surface waves, see \eq{Jconserv}. 

To understand the general structure behind this method, we first present it for a general differential equation of degree $N$, and then apply it to the surface wave equation \eqref{EoM} (App.~\ref{N4_App}). We consider the differential equation 
\be \label{Nmode_eq}
(-i\p_x)^N \phi(x) - \sum_{n=0}^{N-1} f_n(x) (-i\p_x)^n \phi(x) - i \sum_{n=0}^{N-1} g_n(x) (-i\p_x)^n \phi(x) = 0.
\ee
Here, the functions $f_n(x)$ are assumed to be \emph{real} while the $g_n(x)$ can be complex. To this equation, we associate the corresponding Hamilton-Jacobi equation 
\be \label{NHJ_eq}
P_{\rm HJ}(k) = k^N - \sum_{n=0}^{N-1} f_n(x) k^n = 0, 
\ee
where we introduced the Hamilton-Jacobi polynomial $P_{\rm HJ}$. We see that the $g_n$'s do not appear in the Hamilton-Jacobi equation. The reason is that while the $f_n$'s  represent the background as perceived by the field $\phi(x)$, the $g_n$'s represent the features of the wave equation that are absent of the Hamilton-Jacobi equation. In other words, they encode the possible orderings one can choose when promoting $k$ in \eqref{NHJ_eq} as the operator $-i \p_x$ to obtain \eqref{Nmode_eq}. Hence, in the method we shall present, we treat the $g_n$'s as small quantities, as the same order as the \emph{gradients} of the background, i.e. the $f_n'$'s. Going back to \eqref{NHJ_eq}, since $P_{\rm HJ}$ is a polynomial of degree $N$, it has $N$ different roots. The key assumption of the following derivation, is that for all $x$, the $N$ roots are \emph{real} and \emph{distinct}. In particular, no crossing, where one would have $k_j(x) = k_{\ell \neq j}(x)$ for some $x$, occurs. A common procedure with higher order ODEs is to trade the scalar equation of degree $N$ \eqref{Nmode_eq} for a vectorial equation of degree 1. For this, we gather $\phi$ and its derivatives in a column vector
\be
\Phi = \bmat \phi(x) \\ -i\p_x \phi(x) \\ \vdots \\ (-i\p_x)^{N-1}\phi(x) \emat, 
\ee
where the $-i$'s are here for future convenience. \eq{Nmode_eq} then takes the simple matricial form 
\be
-i \p_x \Phi = C(x) \cdot \Phi(x) + i D(x) \cdot \Phi(x), 
\ee
where 
\be \label{CDmat}
C(x) = \bmat 0 & 1 & & (0) \\ \vdots & \ddots & \ddots & \\ 0 & \dots & 0 & 1 \\ f_0 & f_1 & \dots & f_{N-1} \emat \qquad \textrm{and} \qquad D(x) = \bmat 0 & 0 & \dots & 0 \\ \vdots & \vdots &  & \vdots \\ 0 & 0 & \dots & 0 \\ g_0 & g_1 & \dots & g_{N-1} \emat. 
\ee 
$C(x)$ is the $N\times N$ companion matrix associated with the polynomial $P_{\rm HJ}$. The key idea of the Bremmer approach is to ``locally diagonalize $C(x)$'', i.e. at fixed $x$, and then use the eigen-basis to rewrite the original equation \eqref{Nmode_eq}. The characteristic polynomial of $C$ is simply $P_{\rm HJ}$, and therefore the roots of the Hamilton-Jacobi equation are the eigen-values of $C$. Hence, we write our field, solution of \eqref{Nmode_eq}, as 
\be
\phi(x) = \sum_{j=1}^N A_j(x) e^{i S_j(x)}, 
\ee
where the phases are the primitive integral of the Hamilton-Jacobi roots, i.e. 
\be
S_j(x) = \int k_j(x') dx'. 
\ee
Since this introduces $N$ new unknown functions $(A_j)_{j=1..N}$ instead of one, we further impose a similar relation between all derivatives of $\phi$ in $\Phi$ and the local scattering coefficients $A_j$. We define the $A_j$ such that 
\be \label{loc_scat}
\Phi(x) = V \cdot \bmat A_j(x) e^{i S_j(x)} \emat_{j=1..N}, 
\ee
where $V$ is the Vandermonde matrix of the $N$ roots $k_j$, i.e. 
\be
V = \bmat 1 & \dots & 1 \\ k_1 & \dots & k_N \\ \vdots & & \vdots \\ k_1^{N-1} & \dots & k_N^{N-1} \emat. 
\ee
Since all the roots are distinct, $\det(V) = \prod_{j<i} (k_i - k_j) \neq 0$, and the correspondance \eqref{loc_scat} between $\Phi$ and the $A_j$'s is one-to-one. Physically, \eq{loc_scat} means that the $N-1$ first derivatives of $\phi$ act as if the $A_j(x)$ were constant. We cannot impose this to the $N$-th derivative, since the field must be a solution of the wave equation \eqref{Nmode_eq}. This last condition will instead give us a differential equation satisfied by the $A_j(x)$. To obtain it, we first notice that the Vandermonde matrix $V$ is the diagonalizing matrix of the companion matrix $C$. Indeed, it is rather easy to check that 
\be
C \cdot V = V \cdot \mathrm{diag}(k_1, \dots, k_N). 
\ee
We now have enough material to rewrite the mode equation \eqref{Nmode_eq} in a simple manner. For this, we start by deriving the definition of the local scattering coefficients \eqref{loc_scat} 
\bea
-i \p_x \Phi &=& -i\p_x V \cdot \bmat A_j e^{iS_j} \emat + V \cdot \mathrm{diag}(k_1, \dots, k_N) \cdot \bmat A_j e^{iS_j} \emat -i V \cdot \bmat \p_x A_j e^{iS_j} \emat , \nonumber \\
(C + i D) \cdot \Phi &=& -i\p_x V \cdot \bmat A_j e^{iS_j} \emat + C \cdot V \cdot \bmat A_j e^{iS_j} \emat - i V \cdot \bmat \p_x A_j e^{iS_j} \emat , \nonumber \\
i D \cdot \Phi &=& -i\p_x V \cdot \bmat A_j e^{iS_j} \emat - i V \cdot \bmat \p_x A_j e^{iS_j} \emat , \nonumber \\
\bmat \p_x A_j e^{iS_j} \emat &=& - \big[ V^{-1} \p_x V + V^{-1} D V \big] \cdot \bmat A_j e^{iS_j} \emat. \label{unnorm_scat_eq}
\eea
This form is still not well suited for a perturbative resolution. What we want is to partially integrate this equation by normalizing the local scattering coefficients. We define the normalized coefficients $A_j = \mathcal{N}_j a_j$. The prefactors $\mathcal{N}_j$ are chosen in order to get rid of the diagonal elements in the matrix of \eq{unnorm_scat_eq}. This will recast the equation governing the local scattering coefficients into 
\be
\bmat \p_x a_j e^{iS_j} \emat = \mathcal M \cdot \bmat a_j e^{iS_j} \emat , 
\ee
where the diagonal elements of $\mathcal M$ are 0. This guarantees that at 0-th order, one recovers the WKB approximation, i.e. the local scattering coefficients are constant. Inserting $A_j = \mathcal{N}_j a_j$ in \eq{unnorm_scat_eq}, we get the condition for the prefactors 
\be \label{Nampl}
\frac{\p_x \mathcal{N}_j}{\mathcal{N}_j} = -[V^{-1} \p_x V + V^{-1} D V]_{jj}. 
\ee
The matrix element $[V^{-1} \p_x V + V^{-1} D V]_{jj}$ is real and nonsingular, which guarantees that $\mathcal{N}_j$ stays real and positive. After integrating this equation, we finally deduce the equation governing the local scattering coefficients
\be \label{localScatDyn}
\p_x a_\ell = \sum_{j \neq \ell} \mathcal M_{\ell j} e^{i (S_j - S_\ell)} a_j.
\ee
with 
\be \label{brutMij}
[\mathcal M]_{\ell j} = - [V^{-1} \p_x V + V^{-1} D V]_{\ell j} \frac{\mathcal{N}_j}{\mathcal{N}_\ell}. 
\ee

\subsection{Computation of the $\mathcal M$-matrix elements and prefactors $\mathcal{N}_j$}
\label{Mmat_App}
To analyze the various coefficients of the $\mathcal M$-matrix, we will first explicitly integrate the equation for the prefactors $\mathcal{N}_k$ in \eqref{Nampl}. For this we first need to compute the matrix $V^{-1}$. To do so, we introduce $N$ reduced polynomials $P_j$ such that $P_{\rm HJ}(k) = (k - k_j) P_j(k)$, i.e. 
\bea
P_j(k) = \prod_{\ell \neq j} (k - k_\ell) = \sum_{\ell =1}^{N} \alpha_\ell^{\phantom{\ell} j} k^{\ell -1}. 
\eea 
Using them, the coefficients of $V^{-1}$ are easily expressed as 
\be \label{invVdM}
[V^{-1}]_{\ell j} = \frac{\alpha_j^{\phantom{\ell} \ell}}{P_\ell(k_\ell)}. 
\ee
The coefficients $\alpha_\ell^j$ are symmetric polynomials of the $N-1$ roots $k_{\ell \neq j}$. Their exact expression is rather involved, however, for our present purpose, it is not necessary to write them explicitly. Indeed, using \eqref{invVdM}, a direct computation gives 
\be \label{VpxVbrut}
[V^{-1} \p_x V]_{\ell j} = \frac{P_\ell'(k_j) \p_x k_j}{P_\ell(k_\ell)}, 
\ee
where the $'$ denotes derivative with respect to $k$. Moreover, since $P_{\rm HJ}(k) = (k - k_j) P_j(k)$, the derivatives of $P_j$ can be expressed as derivatives of $P_{\rm HJ}$. In particular, the diagonal term of $V^{-1} \p_x V$ reads 
\be
[V^{-1} \p_x V]_{jj} = \frac{P_{\rm HJ}''(k_j) \p_x k_j}{2P_{\rm HJ}'(k_j)}. 
\ee 
At this level, it is tempting to identify the right-hand side of this equation as the logarithmic derivative of $P_{\rm HJ}'(k_j)$. However, one should not forget that the expression $P_{\rm HJ}'(k_j)$ depends on $x$ through both the root $k_j$ and the coefficients of $P_{\rm HJ}'(k)$, i.e. the $f_n(x)$ of \eq{NHJ_eq}. To go further, we must take into account the contribution of the $D$-matrix of \eq{CDmat}. To easily express the coefficients of $V^{-1} D V$, we introduce the new polynomial 
\be \label{Qpol}
Q(k) = \sum_{j=0}^{N-1} g_j(x) k^j. 
\ee
Using it, the combination of \eqref{invVdM} and \eqref{CDmat} shows that 
\be \label{brutDmat}
[V^{-1} D V]_{\ell j} = \frac{Q(k_j)}{P_{\rm HJ}'(k_\ell)} . 
\ee
Therefore, the prefactor equation \eqref{Nampl} rewrites 
\be 
\frac{\p_x \mathcal{N}_j}{\mathcal{N}_j} = -\frac{P_{\rm HJ}''(k_j) \p_x k_j + 2Q(k_j)}{2P_{\rm HJ}'(k_j)} . 
\ee
If the polynomial $Q$ has the good form, this equation directly integrates. Moreover, we recall that unlike $P_{\rm HJ}$, $Q$ might be complex. While its real part contributes to the amplitude of the prefactor, its imaginary part generates a phase. We call the equation \eqref{Nmode_eq} ``$\lam$-canonical'', if there exists a function $\lam(x) > 0$ such that 
\be \label{lamcan}
2 \Re [Q(k)] = \p_x P_{\rm HJ}'(k) + \frac{\lam'(x)}{\lam(x)} P_{\rm HJ}'(k). 
\ee
Under this assumption, the prefactor equation \eqref{Nampl} directly integrates, and one finds 
\be \label{brutpref}
\mathcal{N}_j = \left|\lam(x) P_{\rm HJ}'(k_j) \right|^{-1/2} e^{-i \int \Im[Q(k_j)]/P_{\rm HJ}'(k_j) dx}. 
\ee 
By construction, this generalizes the adiabatic invariant of Sec.~\ref{WKB_Sec}. The phase shift that appears in \eqref{brutpref} might directly affect the scattering coefficients, e.g., by altering the resonance condition \eqref{Eq_sp}. However, in our case this term stays negligible. We now turn to the computation of the matrix elements $\mathcal M_{\ell j}(x)$. For this, we first express \eq{VpxVbrut} in terms of the Hamilton-Jacobi polynomial, and obtain 
\be \label{jell_matel}
[V^{-1} \p_x V]_{\ell \neq j} = -\frac{P_{\rm HJ}'(k_j) \p_x k_j}{(k_j - k_\ell)P_{\rm HJ}'(k_\ell)} . 
\ee
Combining this with the $D$-matrix, the prefactor $\mathcal N$, i.e. Eqs.~\eqref{brutDmat} and~\eqref{brutpref}, the expression \eqref{brutMij} becomes 
\be \label{Mij}
\mathcal M_{\ell j}(x) = \frac{\eps_\ell P_{\rm HJ}'(k_j) \p_x k_j - \eps_\ell (k_j - k_\ell) Q(k_j)}{(k_j - k_\ell) \sqrt{\left| P_{\rm HJ}'(k_\ell) P_{\rm HJ}'(k_j) \right|}}, 
\ee
where $\eps_\ell = \sign(P_{\rm HJ}'(k_\ell))$. We have also dropped the phase proportional to $\Im(Q)$ in \eqref{brutpref} for more clarity, but it is straightforward to add it up.

\subsection{Scattering coefficients in the smooth limit}
\label{prefact_App}
\begin{figure}[!ht]
\begin{center}
\includegraphics[width=0.8\columnwidth]{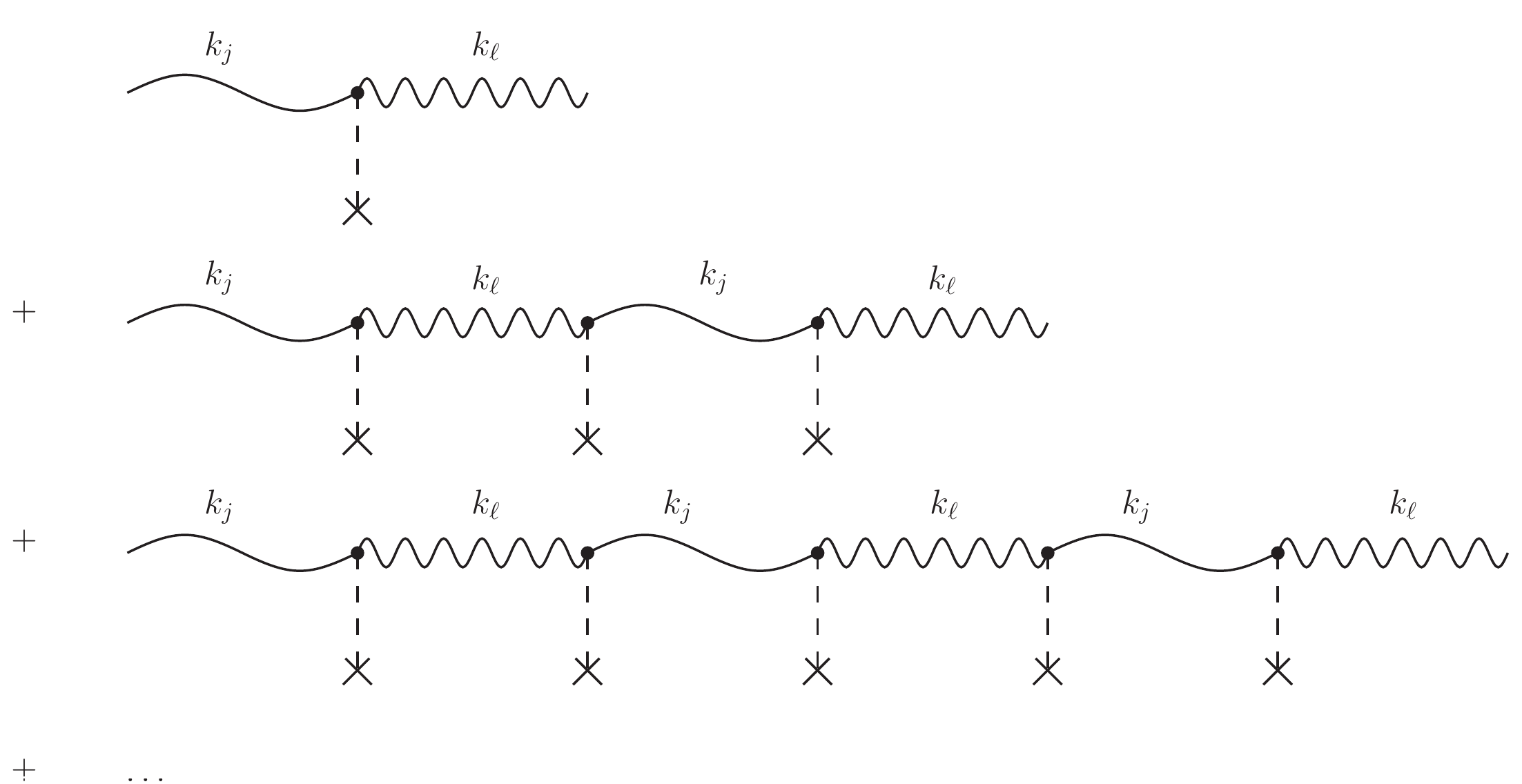}
\end{center}
\caption{We conjecture that the first order expression of the scattering coefficients \eqref{jell_result} can be improved by the replacement $\pi/3 \to 1$. The conjecture is that this replacement consists in resumming the leading contribution in the smooth limit of the diagrams that involve only the roots $k_\ell$ and $k_j$. Once this is done, the higher order perturbative expressions are obtained by summing only over diagrams that involve at least an intermediate state that differs from the initial and final ones. 
}
\label{GraphCpref_Fig} 
\end{figure}

At this level, we point out that the equation governing the local scattering coefficients, i.e. \eqref{localScatDyn}, is fully equivalent to the initial wave equation \eqref{Nmode_eq}. No approximation has been made so far. However, \eqref{localScatDyn} gives a simple perturbative method to compute the scattering coefficients in the limit of slowly varying backgrounds. At first order, \eqref{localScatDyn} shows that the scattering coefficient describing the transition from the mode $j$ to $\ell$ is given by 
\be \label{jell_transition}
\alpha_{j \to \ell} = \int_{-\infty}^{+\infty} \mathcal M_{\ell j}(x) e^{i \int (k_j - k_\ell) dx'} dx. 
\ee
Interestingly, in the smooth limit, that is when $f_n' \to 0$ and $g_n \to 0$, $\alpha_{j \to \ell}$ has a universal behavior. To compute the integral \eqref{jell_transition}, we make a change of variable $S = \int (k_j - k_\ell) dx'$. Since no crossing occurs on the real line (by assumption), $\p_x S = k_j(x) - k_\ell(x) \neq 0$, and our change of variable is licit. \eq{jell_transition} becomes 
\be \label{jell_Cintegral}
\alpha_{j \to \ell} = \int_{-\infty}^{+\infty} \frac{\mathcal M_{\ell j}}{k_j - k_\ell}[S] e^{i S} dS. 
\ee
This integral can now be evaluated by a residue theorem. All we need to do is to pick up the contribution of all the singularities of the integrand such that $\Im(S) > 0$. $\alpha_{j \to \ell}$ is then given by 
\be \label{AlphaSum}
\alpha_{j \to \ell} = 2i\pi \sum_{S_* \in \, {\rm poles}} \mathrm{Res}\left(\frac{\mathcal M_{\ell j}}{k_j - k_\ell}[S];S_*\right) e^{iS_*}. 
\ee
The main type of singularity would be a zero of $k_j - k_\ell$, which is exactly a saddle point $S_* = S(x_*)$. Another type of singularity could arise if one of the background functions $f_n(x)$ has a pole in the complex plane. This second type of singularity usually gives subdominant contribution, as is the case in all the profiles considered in this paper. We shall thus consider the contributions of saddle points only. For this, one should obtain the corresponding residue. Close to the saddle point $x\sim x_*$, we have $k_\ell \sim k_*- \delta k(x)$, and $k_j \sim k_* + \delta k(x)$, and hence 
\be \label{tp_matel}
\mathcal M_{\ell j} \sim \frac{\p_x \delta k}{2 \delta k}. 
\ee
To obtain this equation, we have neglected the contribution of the $Q$-term in \eq{Mij}, but a (rather nontrivial) computation shows that it produces subdominant corrections in $O(\grad)$. Then, using the fact that $\delta k \propto (x - x_*)^{1/2}$ near the saddle point $x_*$, a little algebra shows that 
\be
\frac{\mathcal M_{\ell j}}{k_j - k_\ell}[S] \sim \frac{1}{6(S - S_*)}. 
\ee
Using this, the residue theorem gives us the integral of \eq{jell_transition}, 
\be \label{jell_result}
\alpha_{j \to \ell} = \sum_{S_* \in \, {\rm poles}} \frac{i \pi}{3} e^{i S_*}. 
\ee
As discussed in the text, in this sum, only the terms with the smallest $\Im(S_*)$ contribute significantly, while the other gives exponentially small corrections. Hence the scattering coefficients are usually given by a few contributions, coming from the complex turning points the closest to the real line. We point out here the close similarity between our derivation and the first order result of the Bremmer series~\cite{Berry72,Berry82} (in particular \eq{tp_matel}). In the latter case, $\pi/3$ is only the first term of a series, and in the adiabatic limit, it is possible to show that the entire sum is 1. Therefore, the adiabatic limit leads to $\alpha_{j \to \ell} \sim e^{i S_*}$ rather than \eq{jell_result}. This is also what happens in the adiabatic limit of a time-dependent two-level system~\cite{Davis76}. For this reason, it is reasonable to conjecture that this will also be the case here. If this conjecture holds, the replacement $\pi/3 \to 1$ amounts to a partial resummation of some diagrams of the perturbative resolution of \eq{localScatDyn}, as shown in Fig.~\ref{GraphCpref_Fig}. 

\section{Application to the surface wave equation ($N = 4$)}
\label{N4_App}
\subsection{The Bremmer series for surface waves}
To apply the preceding results to the problem at hand, we must recast \eq{EoM} under the form of \eq{Nmode_eq}. 
\bea
0 &=& \om^2 \phi + 2 \om v i \p_x \phi - (v^2 - c^2) \p_x^2 \phi + \frac{g\h^3}3 \p_x^4 \phi  \nonumber \\
&+& i \om v' \phi + 2 i (vv' - cc') i \p_x \phi + \frac g3 (3 \h^2 \h'' + 6 \h \h'^2) \p_x^2 \phi + 2 g \h^2 \h' \p_x^3 \phi. 
\eea
From this we directly extract the functions $f_n(x)$ and $g_n(x)$. 
\be \label{fg_surf}
\bal
f_0(x) &= -\frac{3 \om^2}{g \h^3} ,  \\
f_1(x) &= \frac{6\om v}{g \h^3} ,  \\
f_2(x) &= \frac{3(c^2 - v^2)}{g \h^3}, \\
f_3(x) &= 0, 
\eal
\qquad \textrm{and} \qquad 
\bal
g_0(x) &= -\frac{3\om v'}{g \h^3} ,  \\
g_1(x) &= \frac{6(vv' - cc')}{g \h^3},  \\
g_2(x) &= -i \frac{(\h^3)''}{\h^3}, \\
g_3(x) &= \frac{6 \h'}{\h} .
\eal
\ee 
Using \eq{lamcan}, we verify that our equation is $\lam$-canonical with the function 
\be \label{lamEq}
\lam(x) = \frac{g \h^3}{6}. 
\ee
Moreover, the quantity $P_{\rm HJ}'(k_j)$ is directly related to the group velocity $v_g^j = (\p_\om k_j)^{-1}$. Indeed, deriving the equation $P_{\rm HJ}(k_j) = 0$ with respect to $\om$ gives us 
\be \label{P'Eq}
P_{\rm HJ}'(k_j) \p_\om k_j + \p_\om P_{\rm HJ} = 0. 
\ee
We also check that the phase shift due to $\Im(Q) \neq 0$ is negligible in the regime of interest, i.e. when $\grad \ll 1$ is valid. Hence, gathering the results of \eqref{lamEq} and \eqref{P'Eq}, we finally obtain 
\be
\mathcal{N}_j = \left|\Om(k_j) v_g(k_j) \right|^{-1/2} ,  
\ee 
where $\Om$ is defined after \eq{DispRelFull}. The expression for the $\mathcal M$-matrix is 
\be
\mathcal M_{\ell j}(x) = \frac{\eps_\ell \Om(k_j) v_g(k_j) \p_x k_j - \eps_\ell g \h^3 (k_j - k_\ell) Q(k_j)/6}{(k_j - k_\ell) \sqrt{\left|\Om(k_j) v_g(k_j) \Om(k_\ell) v_g(k_\ell)\right|}}, 
\ee
with $\eps_\ell = \sign(\Om(k_\ell) v_g(k_\ell))$. From \eq{fg_surf}, we also obtain the expression for $Q$ defined in \eqref{Qpol}, which reads 
\be
\frac{g \h^3}6 Q(k) = - \frac{\om v'}2 + (vv' - cc')k + g \h^2 \h' k^3 - i \frac{g}{6} (\h^3)'' k^2.
\ee
To have an estimation of the magnitude of these matrix elements, we compute them in the limit $\om \ll \omin$. The matrix elements governing the transition $k_\s \to k_\+$ or $k_\s \to k_\-$ are given by  
\be \label{M+u0}
\mathcal M_{\+ \s} \sim - \mathcal M_{\- \s}\sim \frac{\h^{1/2} \om^{1/2} \left[(c^2-v^2)' +(c-v)v'\right]}{2(24)^{1/4} c^{1/4} v^{1/2} (c-v)^{7/4}}. 
\ee
On the other hand, the matrix element driving the transition $k_\s \to k_\co$ reads 
\be \label{Mdu0}
\mathcal M_{\co \s} \sim - \frac{c'}{c}. 
\ee
As we see, both are proportional to derivatives of the background, and are small in the smooth limit. But the matrix elements that govern $\alpha$ and $\beta$ are further suppressed by a factor $O(\om^{1/2})$ at low frequencies, as we mentioned at the end of Sec.~\ref{tp_Sec}. 

\subsection{Scattering coefficients at ultra low frequencies}
\label{oms_App}
To obtain the ultra low frequency behavior of the scattering coefficients, we take the limit $\om \to 0$ \emph{before} the smooth limit $\grad \to 0$. Indeed, as mentioned in Sec.~\ref{tp_Sec}, these two limits do not commute, and the computation of App.~\ref{prefact_App} is valid for $\grad \to 0$ at fixed $\om$. To correctly obtain the $\om \to 0$ limit, we directly use the low frequency expressions of the matrix elements \eqref{M+u0} in the integral representation for $\alpha$ of \eq{AlphaInt} (the computation for $\beta$ gives the same answer in that limit). The coefficient $\alpha$ is then given by 
\be \label{AlphaIntIRfull}
\alpha_\om \sim \int \frac{\h^{1/2} \om^{1/2} \left[(c^2-v^2)' +(c-v)v'\right]}{2(24)^{1/4} c^{1/4} v^{1/2} (c-v)^{7/4}} e^{i \int^x (k_\s(x') - k_\+(x'))dx'} dx, 
\ee
As argued in Sec.~\ref{ShortObst0_Sec}, in near critical flows $1-\Fm \ll 1$, the variations of $1-F(x)$ dominate with respect to the other background quantities. Therefore, at leading order in $1-\Fm$, we have 
\be \label{AlphaIntIR}
\alpha_\om \sim \left(\frac{h_\m \om}{2\sqrt{6} c_\m}\right)^{1/2} \int \frac{(1-F)'}{(1-F)^{7/4}} e^{i \int^x (k_\s(x') - k_\+(x'))dx'} dx, 
\ee
Since we look at the zero frequency limit, in particular we have $\om \ll \omin$, and hence, the phase of the integrant is evaluated using the zero frequency expressions of the roots, i.e. \eq{kZ_al}. Hence the saddle point of \eq{AlphaIntIR} is the complex horizon defined by \eq{Chor}. However, one cannot use a residue theorem to compute the integral as was done in Sec.~\ref{prefact_App}. The reason is that if we make the same change of variable, the point $S_* = \int^{\xctp} (k_\s(x') - k_\+(x'))dx'$ is a branch point and not a pole. Instead, we compute \eqref{AlphaIntIR} using a saddle point theorem~\cite{Olver}. This gives 
\be
\alpha_\om \sim e^{i\frac{3\pi}{4}} \frac{2}{3} \left(\frac{4 \pi \om}{3 c_\m (-F'_*)}\right)^{1/2} e^{i S_*}. 
\ee
From this we deduce 
\be \label{AlphaIR}
|\alpha_\om|^2 \sim \frac{32\sqrt{2} \pi (1-\Fm)^{3/2}}{27 h_\m |F'_*|} \frac{\om}{\omin} e^{- 2 \Im(S_*)}. 
\ee
We see that in the ultra low frequency limit, $\alpha$ is still governed by the same exponential involving the complex turning point (complex horizon in this regime) as in \eq{jell_result}, but the prefactor is not 1. Instead, the prefactor vanishes as $|C|^2 \sim \om/\oms$. The characteristic frequency $\oms$ is defined from \eq{AlphaIR} and reads 
\be \label{oms}
\oms = \frac{27 h_\m |F'_*|}{32\sqrt{2} \pi (1-\Fm)^{3/2}} \omin. 
\ee
We see that $\oms/\omin$ is proportional to a derivative of the background, and therefore is small in the smooth limit. To obtain the simple expression \eqref{oms}, we applied a saddle point approximation, which is valid if~\cite{Olver}
\be
\left|\frac{h_\m^{4/3} (F''_*)^{2}}{(-F'_*)^{8/3}}\right| \ll 1. 
\ee
The above calculations and that of App.~\ref{prefact_App} show that the prefactor is mainly characterized by two behaviors. When $\om \ll \oms$, it vanishes as $\om/\oms$, but for $\oms \ll \om \lesssim \omin$, it approaches 1. To reproduce this behavior, we conjectured an explicit form of the prefactor as a function of $\om$ and used it for the numerical plots of Figs.~\ref{AlphaBeta_Fig}, \ref{Ratio_Fig}, and \ref{2StepAlphaBeta_Fig}. This form is 
\be \label{Ceff}
|C|^2 \sim \left(1-e^{-{|\om|/\oms}}\right) . 
\ee
The absolute value on $\om$ is here to remind us that the prefactor is the same for $\alpha_\om$ and $\beta_\om$, and does not vary when one applies the relation \eqref{AlBetSym}. To conclude this subsection, we wish to underline the fact that this prefactor is an effective way of describing the ultra low frequency behavior of the scattering coefficients. Indeed, were we able to sum over \emph{all} the singularities of \eq{AlphaSum}, we would presumably obtain an expression valid for all values of $\om \lesssim \omin$, including the limit $\om \ll \oms$. It is when we restrict the sum to its dominant contribution that we lose the possibility of taking the limit $\om \to 0$. The ``effective'' prefactor described in Eqs.~\eqref{AlphaIR} and \eqref{Ceff} allows us to recover the correct ultra low frequency behavior.

\subsection{The conserved current}
\label{Current_App}
In this subsection, we briefly sketch how the conserved current $J$ of \eq{Current_Eq} is obtained and how it applies to a superposition of plane waves. This conserved current can be directly obtained by applying the Noether theorem to the action \eqref{action} with the symmetry $\phi \to e^{i \lam} \phi$. A slightly quicker way is to start from the conserved norm of \eq{KGnorm}. Indeed, the Noether theorem says that the norm density $\rho = - \Im \Big(\phi^* (\p_t + v \p_x) \phi \Big)$ and the current $J$ are related by the conservation law
\be
\p_t \rho + \p_x J = 0. 
\ee
By computing $\p_t \rho$ and using the wave equation \eqref{TEoM}, we deduce the current. The calculation is as follows: 
\bsub \bea
\p_t \rho &=& - \Im \Big(\phi^* \p_t^2 \phi + v \p_t \phi^* \p_x \phi + v \phi^* \p_t \p_x \phi \Big) , \label{1} \\
&=& - \Im \Big( v \p_t \phi^* \p_x \phi - \phi^* \p_t \p_x \phi - \phi^* \p_x (v^2 - c^2)\p_x \phi + \frac g3 \phi^* \p_x^2 \h^3 \p_x^2 \phi \Big) , \label{2} \\
&=& - \Im \Big( - v \p_t \phi \p_x \phi^* - \phi^* \p_t \p_x \phi - \phi^* \p_x (v^2 - c^2)\p_x \phi + \frac g3 \phi^* \p_x^2 \h^3 \p_x^2 \phi \Big) , \\
&=& - \Im \Big( \p_x \Big[- v \phi^* \p_t \phi - (v^2 - c^2) \phi^* \p_x \phi \Big] + \underbrace{\p_x \phi^* (v^2 - c^2)\p_x \phi}_{\in \mathbb R} + \frac g3 \phi^* \p_x^2 \h^3 \p_x^2 \phi \Big) , \\
&=& - \Im \Big( \p_x \Big[ - v \phi^* \p_t \phi - (v^2 - c^2) \phi^* \p_x \phi + \frac{g}3 \phi^* \p_x \h^3 \p_x^2 \phi \Big] - \frac{g\h^3}3 \p_x \phi^* \p_x^3 \phi \Big) , \\
&=& - \Im \Big( \p_x \Big[ - v \phi^* \p_t \phi - (v^2 - c^2) \phi^* \p_x \phi + \frac{g}3 \phi^* \p_x \h^3 \p_x^2 \phi - \frac{g\h^3}3 \p_x \phi^* \p_x^2 \phi \Big] \\
&& \qquad \qquad \qquad \qquad \qquad \qquad \qquad \qquad \qquad \qquad \qquad + \underbrace{\frac{g\h^3}3 \p_x^2 \phi^* \p_x^2 \phi}_{\in \mathbb R} \Big) , \\
&=& - \p_x \Im \Big( - v \phi^* \p_t \phi + (c^2 - v^2) \phi^* \p_x \phi + \frac{g}3 \phi^* \p_x \h^3 \p_x^2 \phi - \frac{g\h^3}3 \p_x \phi^* \p_x^2 \phi \Big) , \\
&=& - \p_x J. 
\eea \esub
When we apply $\p_t \rho + \p_x J = 0$ to stationary solutions $\phi(t,x) = \Re(\phi_\om(x) e^{-i \om t})$, we see that the current \eqref{Current_Eq} is $x$-independent. We now want to apply this current to a local superposition of WKB waves as in \eq{Adiab0} so as to obtain \eq{Jconserv}. Because $J$ involves only derivatives of $\phi$ up to third order, the \emph{ansatz} of Eqs.~\eqref{Adiab0}, \eqref{Adiab1}, \eqref{Adiab2}, and \eqref{Adiab3} shows that the computation for a local WKB superposition is the \emph{same} as for exact plane waves, i.e. when $v$, $c$, and $\h$ are constant. Moreover, since $J$ is a quadratic quantity in the field, it is enough to show this for a superposition of 2 plane waves 
\be
\phi = A_1 e^{i k_1 x} + A_2 e^{i k_2 x}, 
\ee
where $k_1$ and $k_2$ are solutions of the dispersion relation \eqref{Disp_rel}. Injecting the above form in \eqref{Current_Eq}, we see that 
\be
J[\phi] = \Om(k_1) v_g(k_1) |A_1|^2 + \Om(k_2) v_g(k_2) |A_2|^2 + J_\times,
\ee
where 
\be \label{Cross_Current}
J_\times = \Big( 2 \om v + (c^2 - v^2)(k_1+k_2) - \frac{g\h^3}{3} \big(k_1^3 + k_2^3 + k_2 k_1^2 + k_1 k_2^2 \big)\Big) \Re \big(A_1^* A_2 e^{-i (k_1-k_2)x} \big). 
\ee
The factor $\Om(k) v_g(k)$ in the diagonal terms directly follows from \eq{P'Eq}. However, it is much more delicate to show that the cross term $J_\times$ is exactly 0. Since $J$ is by construction independent of $x$, it has to vanish. One could presumably stop here, invoking the latter argument, but it would be instructive to understand why the first factor of \eq{Cross_Current} is always 0. It follows from the fact that $k_1$ and $k_2$ are distinct solutions of the dispersion relation \eqref{Disp_rel}. We have proven this identity, but its derivation is rather involved and it is unclear what its physical interpretation is, or how one could generalize it. We present it in the form of the following lemma: 
\bigskip

\noindent {\bf Lemma:} Let $P$ be the polynomial 
\be
P(k) = k^4 + a k^2 + bk + c. 
\ee
If $k_1$ and $k_2$ are two distinct roots of $P$, we have the identity 
\be
b + a(k_1 + k_2) + k_1^3 + k_2^3 + k_1 k_2^2 + k_2 k_1^2 = 0. 
\ee

\begin{proof} Let $k_3$ and $k_4$ be the two other roots of $P$. We call $S$ the left-hand side of the above equation. We notice that 
\bsub \bea
\frac12 \big( P'(k_1) + P'(k_2) \big) &=& \frac12 (k_1 - k_2) \big( (k_1 - k_3)(k_1 - k_4) - (k_2 - k_3)(k_2 - k_4) \big) , \\
&=& 2 (k_1^3 + k_2^3) + a (k_1 + k_2) + b. 
\eea \esub
Hence, 
\be
S = \frac12 \big( P'(k_1) + P'(k_2) \big) - k_1^3 - k_2^3 + k_1 k_2^2 + k_2 k_1^2 .
\ee
Moreover, 
\be
- k_1^3 - k_2^3 + k_1 k_2^2 + k_2 k_1^2 = (k_1 - k_2) (k_2^2 - k_1^2). 
\ee
Therefore, 
\be
S = \frac12 (k_1 - k_2) \big( (k_1 - k_3)(k_1 - k_4) - (k_2 - k_3)(k_2 - k_4) + 2 k_2^2 - 2 k_1^2\big)
\ee
Since $k_1 \neq k_2$, it is enough to show that $\tilde S = 2S/(k_1-k_2) = 0$. Expanding $\tilde S$, it follows that 
\bsub \bea
\tilde S &=& k_1^2 - k_1 k_3 - k_1 k_4 - k_2^2 + k_2 k_3 + k_2 k_4 + 2 k_2^2 - 2 k_1^2 , \\
&=& k_2^2 - k_1^2 - k_1 k_3 - k_1 k_4 + k_2 k_3 + k_2 k_4 , \\
&=& -(k_1 - k_2)(k_1 + k_2) - k_3 (k_1 - k_2) - k_4 (k_1 - k_2) , \\
&=& -(k_1 - k_2)(k_1 + k_2 + k_3 + k_4) ,\\
&=& 0. 
\eea \esub
At the last line, we used the fact that $k_1 + k_2 + k_3 + k_4 = 0$, which comes from the fact that the $k^3$ coefficient of $P$ is zero. 
\end{proof}

\subsection{Contour integral for a monotonic profile}
\label{Contour_App}
\begin{figure}[!ht]
\begin{center}
\includegraphics[width=0.8\columnwidth]{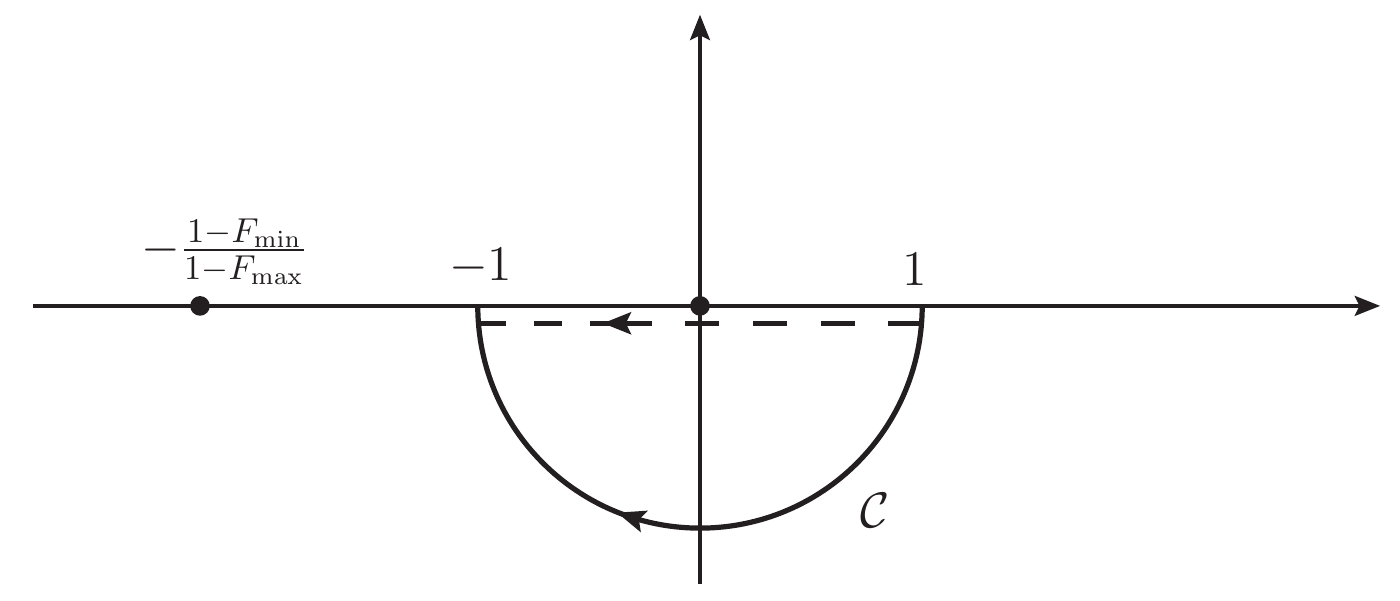}
\end{center}
\caption{Contour of the integral \eqref{1contour}. The bold dots indicate poles of the integrant. The half circle is deformed into the segment $]-1-i \eps; 1-i \eps[$ (dashed line). 
}
\label{1Step_contour_Fig} 
\end{figure}

In this section, we present the computation of the contour integral necessary to obtain the value $\alpha_{\om \to 0}$ in the profile \eqref{1StepProf}, that is, \eq{1StepAlpha}. Interestingly, this computation is very similar to what happens for the Schrödinger equation in a tanh potential~\cite{Berry72}. The integral leading to \eq{1StepAlpha} is defined as 
\bsub \bea
S_* &=& \int_{x_0}^{\xctp} (k_\s(x') - k_\+(x')) dx' , \\ 
&=& \frac{\sqrt{6}}{h_\m} \int_{\xr}^{\xctp} \sqrt{(1 - F(x'))} dx', 
\eea \esub 
where we used $1-\Fm \ll 1$, and we have chosen $x_0 = \xr$ for convenience. Before computing the above integral, we rewrite the single step profile of \eq{1StepProf} as 
\be
F(x) = \Fm - \frac{2D e^{-2\gamma x/D}}{1+e^{-2\gamma x/D}},  
\ee
We compute the contour integral above with the parametrization $x'(t) = \xr + i t \Delta_0$, with $\xr$ and $\Delta_0$ given by \eq{1StepHor}. Using the identity 
\be
e^{-2\gamma x'(t)/D} = \frac{1-F_\m}{1-\Fm} e^{-i \pi t}, 
\ee
we get 
\be \label{B11}
S_* = i \frac{\pi D \sqrt{6}}{2 \gamma h_\m} \int_0^1 \sqrt{1 - \Fm + \frac{2D (1-\Fm) e^{-i \pi t}}{(1-\Fm)e^{-i \pi t}+1-F_\m}} dt. 
\ee
We now interpret this integral as another contour integral, clockwise along the lower half unit circle (see Fig.~\ref{1Step_contour_Fig}). For this we define $z = e^{-i\pi t}$, which implies $dz/z = -i \pi dt$, and rewrite \eqref{B11} as 
\be \label{1contour}
S_* = -\frac{D \sqrt{6(1-\Fm)}}{2 \gamma h_\m} \int_{\mathcal C} \sqrt{1 + \frac{2D z}{(1-\Fm)z + 1-F_\m}} \frac{dz}{z}. 
\ee
We deform this contour into the segment ]-1;1[, right below the pole at $z=0$, which gives
\be
S_* = \frac{D \sqrt{6(1-\Fm)}}{2 \gamma h_\m}  \int_{-1}^1 \sqrt{1 + \frac{2Dz}{(1-\Fm)z + 1-F_\m}} \frac{dz}{z - i\eps}. 
\ee
Finally, using the identity $(z-i\eps)^{-1} = \mathcal P z^{-1} + i \pi \delta(z)$, we get 
\be \label{result_contour}
S_* = i \frac{\sqrt{6} \pi D (1-\Fm)^{1/2}}{2\gamma h_\m}  + \frac{\sqrt{6} D (1-\Fm)^{1/2}}{2 \gamma h_\m}  \mathcal P \int_{-1}^1 \sqrt{1 + \frac{2D}{1-\Fm+(1-F_\m)z}} \frac{dz}{z}. 
\ee
The second term of this equation being real, we have 
\be
\Im(S_*) = \frac{\sqrt{6} \pi D (1-\Fm)^{1/2}}{2\gamma h_\m} ,  
\ee
and hence \eq{1StepAlpha}. We also define $\zeta = \Re(S_*)$, which contributes to the phase shift in \eq{ResPhase}. Using the Cardan-Tartaglia method, and following the steps of Sec.~\ref{OmDep_Sec}, a computation similar to the above one gives us the corrections in $O(\om/\omin)$ of \eq{1StepAlphaom}.

\bibliographystyle{utphys}
\bibliography{Bibli}

\providecommand{\href}[2]{#2}\begingroup\raggedright\begin{thebibliography}{10}

\bibitem{Unruh81}
W.~Unruh, ``{Experimental black hole evaporation},''
\href{http://dx.doi.org/10.1103/PhysRevLett.46.1351}{{\em Phys. Rev. Lett.}
  {\bfseries 46} (1981) 1351--1353}.
%%CITATION = PRLTA,46,1351;%%.

\bibitem{Jacobson91}
T.~Jacobson, ``{Black hole evaporation and ultrashort distances},''
\href{http://dx.doi.org/10.1103/PhysRevD.44.1731}{{\em Phys. Rev.} {\bfseries D
  44} (1991) 1731--1739}.
%%CITATION = PHRVA,D44,1731;%%.

\bibitem{Brout95}
R.~Brout, S.~Massar, R.~Parentani, and P.~Spindel, ``{Hawking radiation without
  transPlanckian frequencies},''
  \href{http://dx.doi.org/10.1103/PhysRevD.52.4559}{{\em Phys. Rev.} {\bfseries
  D 52} (1995) 4559--4568},
\href{http://arxiv.org/abs/hep-th/9506121}{{\ttfamily arXiv:hep-th/9506121
  [hep-th]}}.
%%CITATION = HEP-TH/9506121;%%.

\bibitem{Barcelo05}
C.~Barcelo, S.~Liberati, and M.~Visser, ``{Analogue gravity},'' {\em Living
  Rev. Rel.} {\bfseries 8} (2005) 12,
\href{http://arxiv.org/abs/gr-qc/0505065}{{\ttfamily arXiv:gr-qc/0505065
  [gr-qc]}}.
%%CITATION = GR-QC/0505065;%%.

\bibitem{Coutant14b}
A.~Coutant and R.~Parentani, ``{Hawking radiation with dispersion: The
  broadened horizon paradigm},''
  \href{http://dx.doi.org/10.1103/PhysRevD.90.121501}{{\em Phys. Rev.}
  {\bfseries D 90} (2014) 121501},
\href{http://arxiv.org/abs/1402.2514}{{\ttfamily arXiv:1402.2514 [gr-qc]}}.
%%CITATION = ARXIV:1402.2514;%%.

\bibitem{Steinhauer14}
J.~Steinhauer, ``{Observation of self-amplifying Hawking radiation in an analog
  black hole laser},'' \href{http://dx.doi.org/10.1038/NPHYS3104}{{\em Nature
  Phys.} {\bfseries 10} (2014) 864},
\href{http://arxiv.org/abs/1409.6550}{{\ttfamily arXiv:1409.6550
  [cond-mat.quant-gas]}}.
%%CITATION = ARXIV:1409.6550;%%.

\bibitem{Steinhauer15}
J.~Steinhauer, ``Observation of thermal hawking radiation and its entanglement
  in an analogue black hole,''
  \href{http://arxiv.org/abs/1510.00621}{{\ttfamily 1510.00621}}.
  \url{http://arxiv.org/abs/1510.00621}.

\bibitem{Belgiorno10}
F.~Belgiorno, S.~Cacciatori, M.~Clerici, V.~Gorini, G.~Ortenzi, L.~Rizzi,
  E.~Rubino, V.~Sala, and D.~Faccio, ``Hawking radiation from ultrashort laser
  pulse filaments,''
  \href{http://dx.doi.org/10.1103/PhysRevLett.105.203901}{{\em Phys. Rev.
  Lett.} {\bfseries 105} no.~20, (2010) 203901},
  \href{http://arxiv.org/abs/1009.4634}{{\ttfamily arXiv:1009.4634 [gr-qc]}}.

\bibitem{Rubino12}
E.~Rubino, J.~McLenaghan, S.~C. Kehr, F.~Belgiorno, D.~Townsend, S.~Rohr,
  C.~Kuklewicz, U.~Leonhardt, F.~K{\"o}nig, and D.~Faccio, ``Negative frequency
  resonant radiation,'' {\em Phys. Rev. Lett.} {\bfseries 108} (01, 2012)
  253901, \href{http://arxiv.org/abs/1201.2689}{{\ttfamily 1201.2689}}.
  \url{http://arxiv.org/abs/1201.2689}.

\bibitem{Weinfurtner10}
S.~Weinfurtner, E.~W. Tedford, M.~C. Penrice, W.~G. Unruh, and G.~A. Lawrence,
  ``{Measurement of stimulated Hawking emission in an analogue system},''
  \href{http://dx.doi.org/10.1103/PhysRevLett.106.021302}{{\em Phys. Rev.
  Lett.} {\bfseries 106} (2011) 021302},
\href{http://arxiv.org/abs/1008.1911}{{\ttfamily arXiv:1008.1911 [gr-qc]}}.
%%CITATION = ARXIV:1008.1911;%%.

\bibitem{Weinfurtner13}
S.~Weinfurtner, E.~W. Tedford, M.~C.~J. Penrice, W.~G. Unruh, and G.~A.
  Lawrence, ``{Classical aspects of Hawking radiation verified in analogue
  gravity experiment},''
  \href{http://dx.doi.org/10.1007/978-3-319-00266-8_8}{{\em Lect. Notes Phys.}
  {\bfseries 870} (2013) 167--180},
\href{http://arxiv.org/abs/1302.0375}{{\ttfamily arXiv:1302.0375 [gr-qc]}}.
%%CITATION = ARXIV:1302.0375;%%.

\bibitem{Euve14}
L.-P. Euv{\'e}, F.~Michel, R.~Parentani, and G.~Rousseaux, ``{Wave blocking and
  partial transmission in subcritical flows over an obstacle},''
  \href{http://dx.doi.org/10.1103/PhysRevD.91.024020}{{\em Phys. Rev.}
  {\bfseries D 91} no.~2, (2015) 024020},
\href{http://arxiv.org/abs/1409.3830}{{\ttfamily arXiv:1409.3830 [gr-qc]}}.
%%CITATION = ARXIV:1409.3830;%%.

\bibitem{Euve15}
L.~P. Euv{\'e}, F.~Michel, R.~Parentani, T.~G. Philbin, and G.~Rousseaux,
  ``{Observation of noise correlated by the Hawking effect in a water tank},''
\href{http://arxiv.org/abs/1511.08145}{{\ttfamily arXiv:1511.08145
  [physics.flu-dyn]}}.
%%CITATION = ARXIV:1511.08145;%%.

\bibitem{Volovik06}
G.~Volovik, ``Horizons and ergoregions in superfluids,'' {\em J. LowTemp.
  Phys.} {\bfseries 145} (2006) 337--356,
  \href{http://arxiv.org/abs/gr-qc/0603093}{{\ttfamily arXiv:gr-qc/0603093
  [gr-qc]}}.

\bibitem{Schutzhold02}
R.~Schutzhold and W.~G. Unruh, ``{Gravity wave analogs of black holes},''
  \href{http://dx.doi.org/10.1103/PhysRevD.66.044019}{{\em Phys. Rev.}
  {\bfseries D 66} (2002) 044019},
\href{http://arxiv.org/abs/gr-qc/0205099}{{\ttfamily arXiv:gr-qc/0205099
  [gr-qc]}}.
%%CITATION = GR-QC/0205099;%%.

\bibitem{Unruh12}
W.~G. Unruh, ``{Irrotational, two-dimensional Surface waves in fluids},''
  \href{http://dx.doi.org/10.1007/978-3-319-00266-8_4}{{\em Lect. Notes Phys.}
  {\bfseries 870} (2013) 63--80},
\href{http://arxiv.org/abs/1205.6751}{{\ttfamily arXiv:1205.6751 [gr-qc]}}.
%%CITATION = ARXIV:1205.6751;%%.

\bibitem{Johnson}
R.~Johnson, {\em A modern introduction to the mathematical theory of water
  waves}, vol.~19.
\newblock Cambridge University Press, 1997.

\bibitem{Mei}
C.~C. Mei, {\em The applied dynamics of ocean surface waves}, vol.~1.
\newblock World scientific, 1989.

\bibitem{Coutant12}
A.~Coutant, A.~Fabbri, R.~Parentani, R.~Balbinot, and P.~Anderson, ``{Hawking
  radiation of massive modes and undulations},''
  \href{http://dx.doi.org/10.1103/PhysRevD.86.064022}{{\em Phys. Rev.}
  {\bfseries D 86} (2012) 064022},
\href{http://arxiv.org/abs/1206.2658}{{\ttfamily arXiv:1206.2658 [gr-qc]}}.
%%CITATION = ARXIV:1206.2658;%%.

\bibitem{Coutant13}
A.~Coutant and R.~Parentani, ``{Undulations from amplified low frequency
  surface waves},'' \href{http://dx.doi.org/10.1063/1.4872025}{{\em Phys.
  Fluids} {\bfseries 26} (2014) 044106},
\href{http://arxiv.org/abs/1211.2001}{{\ttfamily arXiv:1211.2001
  [physics.flu-dyn]}}.
%%CITATION = ARXIV:1211.2001;%%.

\bibitem{Finazzi11}
S.~Finazzi and R.~Parentani, ``{On the robustness of acoustic black hole
  spectra},'' {\em J. Phys. Conf. Ser.} {\bfseries 314} (2011) 012030,
\href{http://arxiv.org/abs/1102.1452}{{\ttfamily arXiv:1102.1452 [gr-qc]}}.
%%CITATION = ARXIV:1102.1452;%%.

\bibitem{Michel14}
F.~Michel and R.~Parentani, ``{Probing the thermal character of analogue
  Hawking radiation for shallow water waves?},''
  \href{http://dx.doi.org/10.1103/PhysRevD.90.044033}{{\em Phys. Rev.}
  {\bfseries D 90} no.~4, (2014) 044033},
\href{http://arxiv.org/abs/1404.7482}{{\ttfamily arXiv:1404.7482 [gr-qc]}}.
%%CITATION = ARXIV:1404.7482;%%.

\bibitem{Michel15}
F.~Michel and R.~Parentani, ``{Mode mixing in sub- and trans-critical flows
  over an obstacle: When should Hawking's predictions be recovered?},''
\href{http://arxiv.org/abs/1508.02044}{{\ttfamily arXiv:1508.02044 [gr-qc]}}.
%%CITATION = ARXIV:1508.02044;%%.

\bibitem{Bremmer51}
H.~Bremmer, ``The {WKB} approximation as the first term of a geometric-optical
  series,'' {\em Communications on pure and applied mathematics} {\bfseries 4}
  no.~1, (1951) 105--115.

\bibitem{Landauer51}
R.~Landauer, ``Reflections in one-dimensional wave mechanics,'' {\em Physical
  Review} {\bfseries 82} no.~1, (1951) 80.

\bibitem{Berry72}
M.~V. Berry and K.~Mount, ``Semiclassical approximations in wave mechanics,''
  {\em Reports on Progress in Physics} {\bfseries 35} no.~1, (1972) 315.

\bibitem{Auregan15}
Y.~Aur{\'e}gan, P.~Fromholz, F.~Michel, V.~Pagneux, and R.~Parentani, ``{Slow
  sound in a duct, effective transonic flows, and analog black holes},''
  \href{http://dx.doi.org/10.1103/PhysRevD.92.081503}{{\em Phys. Rev.}
  {\bfseries D92} no.~8, (2015) 081503},
\href{http://arxiv.org/abs/1503.02634}{{\ttfamily arXiv:1503.02634 [gr-qc]}}.
%%CITATION = ARXIV:1503.02634;%%.

\bibitem{Fabrikant}
A.~Fabrikant and I.~Stepanyants, {\em Propagation of waves in shear flows},
  vol.~18.
\newblock World Scientific Publishing Company Incorporated, 1998.

\bibitem{Rousseaux13}
G.~Rousseaux, ``The basics of water waves theory for analogue gravity,''
  \href{http://arxiv.org/abs/1203.3018}{{\ttfamily arXiv:1203.3018
  [physics.flu-dyn]}}.

\bibitem{Richartz12}
M.~Richartz, A.~Prain, S.~Weinfurtner, and S.~Liberati, ``{Superradiant
  scattering of dispersive fields},''
  \href{http://dx.doi.org/10.1088/0264-9381/30/8/085009}{{\em Class. Quant.
  Grav.} {\bfseries 30} (2013) 085009},
\href{http://arxiv.org/abs/1208.3601}{{\ttfamily arXiv:1208.3601 [gr-qc]}}.
%%CITATION = ARXIV:1208.3601;%%.

\bibitem{Coutant11}
A.~Coutant, R.~Parentani, and S.~Finazzi, ``{Black hole radiation with short
  distance dispersion, an analytical S-matrix approach},''
  \href{http://dx.doi.org/10.1103/PhysRevD.85.024021}{{\em Phys. Rev.}
  {\bfseries D 85} (2012) 024021},
\href{http://arxiv.org/abs/1108.1821}{{\ttfamily arXiv:1108.1821 [hep-th]}}.
%%CITATION = ARXIV:1108.1821;%%.

\bibitem{Buhler}
O.~B{\"u}hler, {\em Waves and mean flows}.
\newblock Cambridge University Press, 2014.

\bibitem{Bellman}
N.~Bellman and J.~Vasudevan, {\em Wave propagation: an invariant imbedding
  approach}, vol.~17.
\newblock Springer Netherlands, 2012.

\bibitem{Massar97}
S.~Massar and R.~Parentani, ``{Particle creation and nonadiabatic transitions
  in quantum cosmology},''
  \href{http://dx.doi.org/10.1016/S0550-3213(97)00718-9}{{\em Nucl. Phys.}
  {\bfseries B 513} (1998) 375--401},
\href{http://arxiv.org/abs/gr-qc/9706008}{{\ttfamily arXiv:gr-qc/9706008
  [gr-qc]}}.
%%CITATION = GR-QC/9706008;%%.

\bibitem{Winitzki05}
S.~Winitzki, ``{Cosmological particle production and the precision of the WKB
  approximation},'' \href{http://dx.doi.org/10.1103/PhysRevD.72.104011}{{\em
  Phys. Rev.} {\bfseries D 72} (2005) 104011},
\href{http://arxiv.org/abs/gr-qc/0510001}{{\ttfamily arXiv:gr-qc/0510001
  [gr-qc]}}.
%%CITATION = GR-QC/0510001;%%.

\bibitem{Atkinson60}
F.~Atkinson, ``{Wave propagation and the Bremmer series},'' {\em Journal of
  Mathematical Analysis and Applications} {\bfseries 1} no.~3, (1960) 255--276.

\bibitem{Kay61}
I.~Kay, ``{Some remarks concerning the Bremmer series},'' {\em Journal of
  Mathematical Analysis and Applications} {\bfseries 3} no.~1, (1961) 40--49.

\bibitem{Berry82}
M.~Berry, ``Semiclassically weak reflections above analytic and non-analytic
  potential barriers,'' {\em Journal of Physics A: Mathematical and General}
  {\bfseries 15} no.~12, (1982) 3693.

\bibitem{Davis76}
J.~P. Davis and P.~Pechukas, ``Nonadiabatic transitions induced by a
  time-dependent hamiltonian in the semiclassical/adiabatic limit: The
  two-state case,'' {\em The Journal of Chemical Physics} {\bfseries 64} no.~8,
  (1976) 3129--3137.

\bibitem{Macher09b}
J.~Macher and R.~Parentani, ``{Black hole radiation in Bose-Einstein
  condensates},'' \href{http://dx.doi.org/10.1103/PhysRevA.80.043601}{{\em
  Phys. Rev.} {\bfseries A 80} (2009) 043601},
\href{http://arxiv.org/abs/0905.3634}{{\ttfamily arXiv:0905.3634
  [cond-mat.quant-gas]}}.
%%CITATION = ARXIV:0905.3634;%%.

\bibitem{Balbinot06}
R.~Balbinot, A.~Fabbri, S.~Fagnocchi, and R.~Parentani, ``{Hawking radiation
  from acoustic black holes, short distance and back-reaction effects},'' {\em
  Riv. Nuovo Cim.} {\bfseries 28} (2005) 1--55,
\href{http://arxiv.org/abs/gr-qc/0601079}{{\ttfamily arXiv:gr-qc/0601079
  [gr-qc]}}.
%%CITATION = GR-QC/0601079;%%.

\bibitem{Gottfried}
K.~Gottfried and T.~Yan, {\em Quantum mechanics: fundamentals}.
\newblock Springer Verlag, 2nd~ed., 2003.

\bibitem{Keller62}
H.~B. Keller and J.~B. Keller, ``Exponential-like solutions of systems of
  linear ordinary differential equations,'' {\em Journal of the Society for
  Industrial and Applied Mathematics} {\bfseries 10} no.~2, (1962) 246--259.

\bibitem{Delos72II}
J.~B. Delos and W.~R. Thorson, ``{Semiclassical theory of inelastic collisions.
  II. Momentum-space formulation},'' {\em Phys. Rev.} {\bfseries A 6} no.~2,
  (1972) 720.

\bibitem{Coutant14}
A.~Coutant, ``{Unitary and non-unitary transitions around a cosmological
  bounce},'' \href{http://dx.doi.org/10.1103/PhysRevD.89.123524}{{\em Phys.
  Rev.} {\bfseries D 89} (2014) 123524},
\href{http://arxiv.org/abs/1404.5634}{{\ttfamily arXiv:1404.5634 [gr-qc]}}.
%%CITATION = ARXIV:1404.5634;%%.

\bibitem{Coutant15}
A.~Coutant, ``{Semiclassical momentum representation in quantum cosmology},''
  \href{http://dx.doi.org/10.1103/PhysRevD.93.043520}{{\em Phys. Rev.}
  {\bfseries D 93} no.~4, (2016) 043520},
\href{http://arxiv.org/abs/1504.08156}{{\ttfamily arXiv:1504.08156 [gr-qc]}}.
%%CITATION = ARXIV:1504.08156;%%.

\bibitem{Olver}
F.~Olver, {\em Asymptotics and special functions}, vol.~15.
\newblock Academic Press New York, 1974.

\end{thebibliography}\endgroup

\end{document}